\documentstyle[aps,eqsecnum]{revtex}
\begin{document}
\draft
\title{On-site correlation in valence and core states of ferromagnetic nickel}
\author{F.~Manghi, V.~Bellini}
\address{Istituto Nazionale per la Fisica della Materia and \\
 Dipartimento di Fisica,Universit\`a di Modena, Via Campi 213/a, 
I-41100 Modena, Italy }
\author{C.~Arcangeli}
\address{ Max-Planck-Institut f\"ur Festk\"orperforschung,
Heisenbergstr. 1, D-70569 Stuttgart, Germany}
\maketitle
\begin{abstract}
We present a method which allows to include narrow-band
correlation effects into the description of both valence and core states
and we apply it to the prototypical case of nickel. The results of an
ab-initio band calculation are used as input mean-field eigenstates 
for the calculation of self-energy corrections and spectral
functions according to a three-body scattering solution of a multi-orbital
Hubbard hamiltonian. The calculated quasi-particle spectra show a
remarkable agreement with photoemission data in terms of band width,
exchange splitting, satellite energy position of valence states, spin
polarization of both the main line and the satellite of the {\it 3p} core
level. 
\end{abstract} 

\section{Introduction}
\label{sec-intro}

It is well established that the description of electronic states in narrow
band materials requires improvements over the single particle approximation
with a proper inclusion of on-site Coulomb interaction between localized
electrons \cite{davis}. In these systems the itinerant character of valence
electrons which is clearly shown by the energy k-dispersion observed in
photoemission spectroscopy coexists with strong local electronic
correlation responsible of other observed features such as satellite
structures and band-narrowing effects. The interplay of localization and
itineracy has also been indicated as a possible explanation of the observed
spin polarization of core level spectra through exchange coupling betweeen
localized (core) and itinerant (valence) states \cite{Kakehashi,See}. 

In this paper we present a theoretical description of the valence and core
electron states of nickel according to a method recently developed which
has been designed to treat highly correlated and highly hybridized
systems~\cite{Manghi} including both the itinerant character of band
electrons and the strong localized electron-electron repulsion. This method
allows to include narrow-band correlation effects into a
first-principle band calculation; the single-particle band states are
determined according to the density functional theory in the local density
approximation  (LDA) and the correlation effects are described as a 3-body
scattering (3BS) solution of a multi-orbital Hubbard hamiltonian. This approach 
has
been previously applied to the description of valence and conduction states
of both model systems \cite{Igarashi,Calandra} and realistic materials
\cite{Manghi,Villafiorita,Igarashi2}. 
We present here an extension of the method in order to treat
both valence and core state on the same footing;  this extension is
possible in this scheme since it relies  on
a multi-orbital Hubbard hamiltonian where core and valence states can coexist
and  on the 3BS method which can be applied for any value of
effective on-site electron-electron repulsion. 

As far as the valence states are concerned various methods have been
proposed to augment conventional band theory for the description of the
electronic states of nickel; some of them are based on  perturbative
expansion either in the e-e interaction (GW approach \cite{ferdi}, second-
order solution of the Hubbard hamiltonian \cite{treglia,mednick,Jordan}) 
or in the fluctuations of the electron occupation around LDA
mean-field solution \cite{Steiner}; others apply a t-matrix scheme where
the effect of electron correlation on one electron removal energies from a
partially filled band is described as a  hole-hole interaction
\cite{Liebsch,Manghi-ni}. This method is strictly valid only in the limit
of an  almost filled band (dilute limit) and its application to the case of
nickel has been questioned \cite{treglia}. In order to implement this
approach it is necessary to include also electron-electron scattering
channels and to solve a 3-body scattering problem involving two holes and
one electron. This is the spirit of the 3BS theory we apply here
and which originally has been formulated by Igarashi \cite{Igarashi}. This
approach has recently been applied to the description of valence states of
nickel \cite{fulde-ni} choosing, however,  an approximate form of self-energy
in terms of the antisymmetrized vertex function; in this way the self-energy
turns out to be real, giving rise to peaks in the spectral density of
unphysical zero width. Here we will instead adopt a version of the theory
which avoids this shortcoming and which is based on the explicit solution
of the 3-body scattering equations \cite{Manghi,Calandra,Villafiorita}. 

The interpretation of core level spectroscopies has up to now largely been
based on atomic models which interpret the structures observed in the one
electron removal spectra in terms of multiplet states formed by coupling
the core hole to the unfilled valence shell \cite{Fadley}. This scheme has
in particular been applied to the photoemission spectra from the core
states of transition metals \cite{Carbone1} attributing the
observed characteristic line splitting to intra-atomic exchange interaction
between core and valence electrons of an isolated atom. 
Such an approach has been seriously
questioned since the observed splittings and the energy scale of the
interaction are of the same order of magnitude as the valence band width
\cite{Kakehashi}; a picture which takes into account the itinerant
character of valence electrons seems therefore necessary. 
With the advent of spin polarized spectroscopies such as spin resolved
x-ray photoemission   \cite{See,Carbone} and magnetic circular
dichroism \cite{Thole}  the spin dependence of both the main line 
and the satellites of Ni core level spectra
has been widely investigated \cite{Carbone,Seeprl,Liu,Saitoh} . 
We will show that the 3BS solution
of a multi-band Hubbard hamiltonian, where full details of
the valence band structure are included, can account for 
these spectroscopical features and interpret them in terms of on-site 
interaction between localized (core) and itinerant (valence) states.

The paper is organized as follows: we present in section \ref{sec-hh} the
multi-orbital Hubbard hamiltonian and its relationship to the band hamiltonian
we want to implement; section \ref{sec-selfe} describes the main
characteristics of the method which we use to get an approximate solution of the
Hubbard hamiltonian in terms of self-energy corrections to band eigenstates
and of spectral densities; section \ref{sec-val} and \ref{sec-core}
specialize to the case of valence and core states respectively; the results
and the comparison with experiments are presented in section \ref{sec-res}.

\section{Multi-band Hubbard hamiltonian}
\label{sec-hh}
Band structure eigenvalues are in many cases 
good zero-order approximations to  the 
excitation spectrum of a solid and it seems reasonable to use them as a 
starting point for the inclusion of correlation effects according to the 
Hubbard model;  the implicit assumption is that among all the many body 
terms responsible of electron correlation,
the Coulomb repulsion between electrons on the same site is the one 
which needs to be treated explicitly.
To do this it is necessary to define precisely the 
relationship between band and Hubbard hamiltonians. Let us consider first 
a localized basis set $\phi_{i\alpha\sigma}({\bf r},s)$
with $i$ labelling the localization site, $\alpha$ the orbital character, 
$s$ and $\sigma$ the spin coordinate and eigenvalue respectively.
The full many-body hamiltonian in second quantization is
\[
\hat{H}=
\sum_{i\alpha\sigma} \epsilon^0_{i\alpha\sigma} \hat{n}_{i\alpha\sigma}+
\sum_{\alpha \beta \sigma} \sum_{ij}
t_{i \alpha, j \beta} \hat{c}_{i \alpha \sigma}^{\dagger} 
\hat{c}_{j \beta \sigma}+
\frac{1}{2} \sum_{i \alpha  j \beta l \gamma m \delta}
\sum_{\sigma \sigma^{\prime}}
V_{i \alpha \sigma,j \beta \sigma^{\prime},l\gamma\sigma^{\prime},m\delta
\sigma}\hat{c}_{i\alpha\sigma}^{\dagger}
\hat{c}_{j\beta\sigma^{\prime}}^{\dagger}
\hat{c}_{l\gamma\sigma^{\prime}}
\hat{c}_{m\delta\sigma}
\]
with $\hat{n}_{i\alpha\sigma}=\hat{c}_{i\alpha\sigma}^{\dagger}
\hat{c}_{i\alpha\sigma}$ and 
$\hat{c}_{i\alpha\sigma}, \hat{c}_{i\alpha\sigma}^{\dagger}$
destruction and creation operators.

Here $	\epsilon^0_{i\alpha\sigma}$ and $t_{i\alpha , j\beta}$ are the intra- 
and inter-atomic matrix elements of the one-particle hamiltonian ( kinetic 
energy + ionic potential), while 
$V_{i\alpha\sigma,j\beta\sigma^{\prime},l\gamma\sigma^{\prime},m\delta\sigma}$
are multi-center integrals involving the electron-electron 
interaction
\[
V_{i\alpha\sigma,j\beta\sigma^{\prime},l\gamma\sigma^{\prime},m\delta
\sigma}=\sum_{s s'} \int \phi_{i\alpha\sigma}^{*}({\bf r},s)
\phi_{j\beta\sigma^{\prime}}^{*}({\bf r}^{\prime},s^{\prime})
\frac{e^2}{|{\bf r}-{\bf r}^{\prime}|}
\phi_{l\gamma\sigma^{\prime}}({\bf r}^{\prime},s^{\prime})
\phi_{m\delta\sigma}({\bf r},s) d{\bf r}d{\bf r}^{\prime}
\]
In this last expression the dominant contribution comes from the one-center
integrals with $i=j=l=m$ which are the usual on-site Coulomb term
\[
U^i_{\alpha \beta} = 
V_{i\alpha\sigma,i\beta\sigma,i\beta\sigma,i\alpha\sigma}=
V_{i\alpha\sigma,i\beta -\sigma,i\beta -\sigma,i\alpha\sigma}
\]
and exchange term
\[
J^{i}_{\alpha \beta} = 
V_{i\alpha\sigma,i\beta\sigma,i\alpha\sigma,i\beta\sigma}
\]
The full many body hamiltonian can then be written as
\begin{eqnarray}
\label{acca1}
\hat{H} &=&
\sum_{i\alpha\sigma} \epsilon_{i\alpha\sigma} \hat{n}_{i\alpha\sigma}+
\sum_{\alpha \beta \sigma} \sum_{ij} 
t_{i \alpha, j \beta} \hat{c}_{i \alpha \sigma}^{\dagger} 
\hat{c}_{j \beta \sigma} \nonumber \\
&+&
\frac{1}{2}
\sum_{ \alpha \beta} \left [ \sum_{i} (U^i_{\alpha \beta} - J^i_{\alpha \beta})
\sum_{\sigma} \hat{n}_{i\alpha\sigma} \hat{n}_{i\beta\sigma}+
\sum_{i} U^i_{\alpha \beta} 
\sum_{\sigma} \hat{n}_{i\alpha\sigma} \hat{n}_{i\beta -\sigma}\right ] \nonumber \\
&+&
... \mbox{(multi-center terms)}
\end{eqnarray}
Different approximations of the exact hamiltonian (\ref{acca1}) can be
obtained using a mean field approach which amounts to neglect
fluctuations in the electron occupation 
\begin{eqnarray*}
\hat{n}_{i\alpha\sigma} \hat{n}_{i\beta \sigma'} &=& 
\hat{n}_{i\alpha\sigma} <\hat{n}_{i\beta \sigma'}> +
\hat{n}_{i\beta\sigma'} <\hat{n}_{i\alpha \sigma}> -
<\hat{n}_{i\alpha\sigma}> <\hat{n}_{i\beta \sigma'}> \nonumber \\
&+& 
(\hat{n}_{i\alpha\sigma}-<\hat{n}_{i\alpha \sigma}>) \cdot
(\hat{n}_{i\beta\sigma'}-<\hat{n}_{i\beta \sigma'}>) \nonumber \\
&\simeq&
\hat{n}_{i\alpha\sigma} <\hat{n}_{i\beta \sigma'}> +
\hat{n}_{i\beta\sigma'} <\hat{n}_{i\alpha \sigma}> -
<\hat{n}_{i\alpha\sigma}> <\hat{n}_{i\beta \sigma'}>
\end{eqnarray*}
where $<..>$ means a ground state average.
The mean-field 
approximation can be applied to all the many body terms of (\ref{acca1})
transforming it into a single-particle hamiltonian
\begin{equation}
\label{accamf}
\hat{H}^{MF}=
\sum_{i\alpha\sigma} 
\epsilon^{MF}_{i \alpha \sigma} \hat{n}_{i\alpha\sigma}+
\sum_{\alpha \beta \sigma} \sum_{ij} 
t_{i \alpha, j \beta} \hat{c}_{i \alpha \sigma}^{\dagger} 
\hat{c}_{j \beta \sigma}
\end{equation}
Any band structure calculation, where the interacting system is described
as an effective single-particle problem, corresponds to the self-consistent
solution of $\hat{H}^{MF}$. 
Another possibility is to apply the mean field approximation selectively to 
the multi-center integrals, keeping the full many body character in 
the one-center terms; in this way one gets a generalized Hubbard model
\begin{eqnarray}
\hat{H}^{H} &=&
\sum_{i\alpha\sigma} \epsilon^H_{i\alpha\sigma} \hat{n}_{i\alpha\sigma}+
\sum_{\alpha \beta \sigma} \sum_{ij} 
t_{i \alpha, j \beta} \hat{c}_{i \alpha \sigma}^{\dagger} 
\hat{c}_{j \beta \sigma} \nonumber \\
&+&
\frac{1}{2}
\sum_{ \alpha \beta}\left [ \sum_{i} (U^i_{\alpha \beta} - J^i_{\alpha \beta})
\sum_{\sigma} \hat{n}_{i\alpha\sigma} \hat{n}_{i\beta\sigma}+
\sum_{i} U^i_{\alpha \beta} 
\sum_{\sigma} \hat{n}_{i\alpha\sigma} \hat{n}_{i\beta -\sigma} \right ]
\end{eqnarray}
Since $H^{MF}$ and $H^{H}$ differ for the treatment of the on-site 
correlation - included in $H^{MF}$ as a mean-field and treated as a many 
body term in $H^H$ - it is easy to show that 
\begin{equation}
\label{emf}
\epsilon^{MF}_{i \alpha \sigma} =
\epsilon^{H}_{i\alpha\sigma} +
\sum_{  \beta} \left [ (U^i_{\alpha \beta} - J^i_{\alpha \beta})
<\hat{n}_{i\beta\sigma}> +
U^i_{\alpha \beta} <\hat{n}_{i\beta -\sigma}> \right ] 
\end{equation}

Due to the translational periodicity we can 
introduce an extended Bloch basis set
\begin{equation}
\label{bloch1}
\psi^n_{{\bf k}\sigma}({\bf r},s)=\frac{1}{\sqrt{N}}\sum_{i\alpha}
C^n_{\alpha \sigma}({\bf k})
e^{i{\bf k}\cdot{\bf R}_i}\phi_{i\alpha\sigma}({\bf r},s)
\end{equation}
and the corresponding relations for creation/destruction operators
of electrons with wave vector
${\bf k}$ , spin $\sigma$ and band index $n$
 \[
\hat{a}^n_{{\bf k}\sigma}=
\frac{1}{\sqrt{N}}\sum_{i\alpha}C^n_{\alpha \sigma}
({\bf k})e^{i{\bf k}\cdot{\bf R}_i}\hat{c}_{i\alpha
\sigma} 
 \]
 \[ 
\hat{a}^{n^{\dagger}}_{{\bf k}\sigma}=
\frac{1}{\sqrt{N}}\sum_{i\alpha}{C^n_{\alpha \sigma}}({\bf k})^{*}
e^{-i{\bf k}\cdot{\bf R}_i}
\hat{c}^{\dagger}_{i\alpha\sigma}
 \]
Here $C^n_{\alpha}({\bf k}\sigma)$ are the expansion coefficients of
Bloch states in terms of localized orbitals, $N$ the number of unit cells. 
$H^{H}$ becomes 
\begin{eqnarray}
\label{hhk}
\hat{H}^H  &=&  
\sum_{{\bf k} n \sigma}\epsilon^H_{{\bf k}n\sigma}
\hat{a}^{n^{\dagger}}_{{\bf k}\sigma}\hat{a}^n_{{\bf k}\sigma} 
+ \sum_{\alpha\beta}
\sum_{{\bf k}{\bf k}'{\bf p}}
\sum_{n n'} \sum_{m m'}
\sum_{\sigma}
\frac{1}{2N}\cdot \nonumber \\
& &\cdot \left [U_{\alpha\beta}
C_{\alpha \sigma}^n({\bf k})^{*}
C_{\alpha \sigma}^{n'}({\bf k}+{\bf p})
C_{\beta -\sigma}^m({\bf k}')^{*}
C_{\beta -\sigma}^{m'}({\bf k}'-{\bf p})
\hat{a}^{n\dagger}_{{\bf k} \sigma}
\hat{a}^{ n'}_{{\bf k}+{\bf p} \sigma}
\hat{a}^{m\dagger}_{{\bf k}'-\sigma}
\hat{a}^{m'}_{{\bf k}'-{\bf p} -\sigma}\right. \nonumber \\
&+&
\left. (U_{\alpha\beta}-J_{\alpha\beta})
C_{\alpha \sigma}^n({\bf k})^{*}
C_{\alpha \sigma}^{n'}({\bf k}+{\bf p})
C_{\beta \sigma}^m({\bf k}')^{*}
C_{\beta \sigma}^{m'}({\bf k}'-{\bf p})
\hat{a}^{n\dagger}_{{\bf k} \sigma}
\hat{a}^{ n'}_{{\bf k}+{\bf p} \sigma}
\hat{a}^{m\dagger}_{{\bf k}'\sigma}
\hat{a}^{m'}_{{\bf k}'-{\bf p} \sigma} \right]
\end{eqnarray}

Now $\epsilon^H_{{\bf k}n\sigma}$ includes also the kinetic part of
the single-particle hamiltonian and the Coulomb and exchange integrals are
assumed to be site-independent. In the same way $H^{MF}$ becomes 
\begin{equation}
\label{hbk}
\hat{H}^{MF}  =  
\sum_{{\bf k} n \sigma}\epsilon^{MF}_{{\bf k}n\sigma}
\hat{a}^{n^{\dagger}}_{{\bf k}\sigma}\hat{a}^n_{{\bf k}\sigma}
\end{equation}
with
\begin{equation}
\label{eek}
\epsilon^{MF}_{{\bf k}n\sigma} = \epsilon^{H}_{{\bf k}n\sigma} +
Q^n_{{\bf k} \sigma}
\end{equation}
\begin{equation}
\label{qmf}
Q^n_{{\bf k} \sigma}= \sum_{\alpha \beta} 
|C_{\alpha \sigma}^n({\bf k})|^2 \left[ U_{\alpha \beta}\frac{1}{N} 
\sum^{occ}_{{\bf k'} n'} 
|C_{\beta -\sigma}^{n'}({\bf k'})|^2 + (U_{\alpha \beta}
-J_{\alpha \beta})\frac{1}{N} 
\sum^{occ}_{{\bf k'} n'}|C_{\beta \sigma}^{n'}({\bf k}')|^2 \right] 
\end{equation}

which is the analogue of eq. (\ref{emf}) for Bloch 
states. Notice that the sums over $n'$ are over occupied 
states. 
Equations (\ref{eek},\ref{qmf})
contain the correct recipe to include 
Hubbard correlation starting from band structure eigenvalues 
$\epsilon^{MF}_{{\bf k}n\sigma} $ and are
essential in order to avoid double counting of e-e interaction.

\section{Hole spectral function,  self-energy and the Faddeev method}
\label{sec-selfe}
We are interested in the hole spectral function 
\begin{equation}
\label{specf}
D^{-}_{{\bf k}  \sigma}(\omega) = - \frac{1}{\pi} 
\sum_{n} Im {\cal G}^-\left({\bf k} n \sigma,\omega\right)
\end{equation}
which is the quantity directly related to the photoemission results we want 
to compare with. It describes the removal of 
one electron 
of wave-vector ${\bf k}$, band index $n$ and spin $\sigma$ and is related 
to the hole-propagator
\begin{eqnarray}
\label{GH}
{\cal G}^-\left({\bf k} n \sigma,\omega\right)&=&
-\left<\Psi_0\right|\hat{a}^{n^{\dagger}}_{{\bf k}\sigma}
\hat{G}\left(z\right)
\hat{a}^n_{{\bf k}\sigma} \!\left|\Psi_0\right> \ \ ; \qquad
\qquad z=-\omega+E_0\left(N_e\right)+{\rm i}\delta \nonumber\\
\end{eqnarray}
$E_0\left(N_e\right)$ and $\left|\Psi_0\right>$ define the ground 
state of the $N_e$ particle system and 
\begin{equation}
\hat{G}\left(z\right)=\frac{1}{z-\hat{H}^H}
\label{resolvent}
\end{equation}
is the resolvent operator. 
The hole propagator can also be written in terms of the hole self-energy as 
\begin{eqnarray}   
\label{selfenerg}   
{\cal G}^-\left({\bf k} n \sigma,\omega\right)&=&
\frac{1}{\omega-\epsilon_{{\bf k} n \sigma}^{MF}
-\Sigma^-_{{\bf k} n \sigma} (\omega) }\nonumber\\
\end{eqnarray}
where $\Sigma^-_{{\bf k} n \sigma}(\omega)$ is the self-energy 
correction to {\it band} eigenvalues $\epsilon_{{\bf k} n \sigma}^{MF}$ .
In order to calculate 
$\Sigma^-_{{\bf k} n \sigma}(\omega)$
 we proceed as in ref. \cite{Calandra} 
adopting a configuration-interaction scheme which consists
in projecting the Hubbard hamiltonian on a set of states
obtained by adding a finite number of e-h pairs to the Fermi sea, i.e. to 
the ground state $|\Phi_0>$ of the single-particle hamiltonian. 
We will adopt the 3-body scattering (3BS) approach 
where this expansion is truncated to include just one e-h pair: the 
state whith one removed electron of momentum ${\bf k}$ and spin 
$\sigma$  is expanded in terms of the 
basis set including 1- hole configurations 
and 3-particle configurations (1 hole + 1 e-h pair) we will denote by 
$\left|s\right>$ and $\left|t\right>$ respectively
\begin{equation}
\left|s\right> \equiv \hat{a}_{{\bf k} n \sigma} \left|\Phi_0\right> 
\ \ \ \
\left|t\right> \equiv
\hat{a}^{\dagger}_{{\bf q}_3 n_3 \sigma_3} \hat{a}_{{\bf q}_2 n_2 \sigma_2}
\hat{a}_{{\bf q}_1 n_1 \sigma_1}\left|\Phi_0\right> 
\label{states}
\end{equation}
with 
\[
{\bf q}_1+{\bf q}_2-{\bf q}_3= {\bf k} \ \ ; \qquad
\qquad \sigma_1+\sigma_2-\sigma_3 = \sigma 
\]

To be consistent the basis set for the N-particle interacting system will
include zero- and two-particle configurations. 
The ground state of the interacting $N_e$-particle
system coincides then with the non-interacting one; this is obvious for the
single-band hamiltonian discussed in ref. \cite{Calandra} (the state
with zero and one e-h pair added coincide due to k-vector conservation) and
still holds in the present case of multi-band hamiltonian since $H^H$ 
has no off-diagonal matrix elements among 2-particle configurations.
As a consequence in the 3BS approximation the hole
propagator is the average of the resolvent over states
$\left|s\right>$. 

By projecting the hamiltonian (\ref{hhk}) over the complete set 
appropriate for the $N_e-1$ particle system we get an approximate 
expression for $\hat{H}^H$  appropriate to describe one-electron removal
\begin{equation}
\hat{H}^H_{N_e-1} \simeq \hat{H}_1 + \hat{H}_3 +\hat{V}
\end{equation}
Here $\hat{H}_1$ is associated to one-hole configurations
\[
\hat{H}_1 = \left<s\right|\hat{H}^H\left|s\right> \left|s\right>\left<s\right|
\]
$\hat{H}_3$
describes the contribution of 3-particle configurations 
\[
\hat{H}_3= 
\sum_{t t'} 
\left<t\right|\hat{H}^H\left|t'\right> 
\left|t\right>\left<t'\right|
\]
and $\hat{V}$ is the coupling between 1- and 3-particle states
\[
\hat{V} = \sum_{t} 
\left<s\right|\hat{H}^H\left|t\right> 
\left|s\right>\left<t\right| + h.c.
\]
We leave the detailed expression of the matrix elements of the multi-band
$\hat{H}^H$ to appendix \ref{sec-app1} and proceed to sketch the method for
the calculation of the resolvent (\ref{resolvent}). We define the
3-particle resolvent, that is the resolvent associated to the 3-particle
interaction 
\[
\hat{F}_3(z) =  \frac{1}{z-  \hat{H}_3} 
\]
and the Dyson equation which relates $\hat{G}(z)$ to it
\begin{equation}
\hat{G}(z)=\hat{F}_3(z)+\hat{F}_3(z) [\hat{H}_1+\hat{V}]\hat{G}(z)
\label{dyson}
\end{equation}
It is a matter of simple algebra to show that the hole propagator can then be 
expressed in terms of $\hat{F}_3$ as
\begin{equation}
\label{ftt'}
{\cal G}^-\left({\bf k} n \sigma,\omega\right) =
- G_{ss}(z) = 
\frac{1}{\displaystyle \omega-E_0\left(N_e\right) +H^H_{ss} 
+ \sum_{tt'}F_{3tt'}V_{t's}V_{st}}
\end{equation}
with the notation $ G_{ss} \equiv \left<s\right|\hat{G}\left|s\right>$, 
$ F_{3 tt'}  \equiv 
\left<t\right|\hat{F}_3\left|t'\right> $ etc. 
Since the difference between the ground state energy of the $N_e$-particle 
system $E_0(N_e)$ and the average of $H^H$ over $|s>$ states turns out to 
be
\[
E_0(N_e) - H^H_{ss} = 
\epsilon_{{\bf k} n \sigma}^H + Q^n_{{\bf k} \sigma} = 
\epsilon_{{\bf k} n \sigma}^{MF}
\]
the band eigenvalues appear naturally in the denominator of the hole
propagator and, comparing eq. (\ref{ftt'}) with (\ref{selfenerg}), we can
identify the self-energy correction to band eigenvalues we are
interested in as 
\begin{equation}
\Sigma^-\left({\bf k} n \sigma,\omega\right) =
 - \sum_{tt'}F_{3tt'}V_{t's}V_{st} 
\label{sigmab}
\end{equation}

The determination of 
the hole propagator is then reduced to the 
calculation of $ F_{3tt'}$. This is done according to the Faddeev 
scattering theory \cite{Faddeev} as described for the single band 
case in reference \cite{Calandra}. The method consists in separating the 
3-body hamiltonian in diagonal and non-diagonal parts 
\[
\hat{H}_3^D=
\sum_{t} 
\left<t\right|\hat{H}^H\left|t\right> 
\left|t\right>\left<t\right|
\]
\[
\hat{H}_3^{ND}=
\sum_{t t'} 
\left<t\right|\hat{H}^H\left|t'\right> 
\left|t\right>\left<t'\right|
\]
defining the diagonal 3-body resolvent
\[
\hat{F}_3^D\left(z\right)=\frac{1}{z-\hat{H}_3^D}
\]
and the scattering operator 
\[
\hat{S}=\hat{H}_3^{ND} + \hat{H}_3^{ND} \hat{F}_3^D \hat{S}
\]
The 3-body resolvent can then be written as 
\begin{equation}
\label{f3}
\hat{F}_3 = 
\hat{F}_3^D + 
\hat{F}_3^D   \hat{S}  \hat{F}_3^D
\end{equation}
As shown in Appendix 
\ref{sec-app1} and reference \cite{Calandra} the non-diagonal
3-body interaction is the sum of two potentials 
\[
\hat{H}_3^{ND} = \hat{V}_{h-h} + \hat{V}_{h-e}
\]
which describe h-h and h-e multiple scattering.

We define partial scattering operators 
\begin{eqnarray*}
\hat{S}_{h-h} &=& \hat{V}_{h-h} + \hat{H}_{h-h} \hat{F}_3^D \hat{S} \\ 
\hat{S}_{h-e} &=& \hat{V}_{h-e} + \hat{H}_{h-e} \hat{F}_3^D \hat{S}
\end{eqnarray*}
which are related to the scattering T-matrices 
\begin{mathletters}
\label{tmat}
\begin{equation}
\hat{T}_{h-h} = \hat{V}_{h-h} + \hat{V}_{h-h} \hat{F}_3^D \hat{T}_{h-h} \\
\end{equation}
\begin{equation}
\hat{T}_{h-e} = \hat{V}_{h-e} + \hat{V}_{h-e} \hat{F}_3^D \hat{T}_{h-e}
\end{equation}
\end{mathletters}
through
\begin{mathletters}
\label{Faddeev}
\begin{equation}
\hat{S}_{h-h} = \hat{T}_{h-h} + \hat{T}_{h-h} \hat{F}_3^D \hat{S}_{h-e} 
\end{equation}
\begin{equation}
\hat{S}_{h-e} = \hat{T}_{h-e} + \hat{T}_{h-e} \hat{F}_3^D \hat{S}_{h-h}
\end{equation}
\end{mathletters}
These are the Faddeev equations which must be solved in order to get 
$\hat{S}=\hat{S}_{h-h}+\hat{S}_{h-e}$, $\hat{F}_3$ from (\ref{f3}) and 
${\cal G}^-\left({\bf k} n \sigma,\omega\right)$ from (\ref{ftt'}).
Inserting (\ref{Faddeev}) into  (\ref{f3}) one gets the expression  for the 3-
particle resolvent in terms of scattering operators $S_{h-e}$ and $T_{h-h}$
\begin{equation}
\label{ff3}
\hat{F}_3 = 
\hat{F}_3^D + 
\hat{F}_3^D  \left( \hat{T}_{h-h} + \hat{T}_{h-h}\hat{F}_3^D \hat{S}_{h-e} 
+ S_{h-e} \right)  \hat{F}_3^D
\end{equation}

Further steps are necessary in order to make this general method practical
for real calculations - and further approximations as well. 
In the following we will specialize to valence and core states.

\section{Valence states}
\label{sec-val}
We adopt some semplifying assumptions for the description of
photoemission from valence states of nickel. We
will consider e-e correlation among $d$ electrons only and 
neglect the orbital dependence of one-center integrals involving 
valence $d$ electrons; we put 
\begin{eqnarray*}
U_{\alpha \beta} &=& U_{v v} \,\,\,\,\mbox{for\,\,} \alpha, \beta =\, d 
\mbox{orbitals}
\\ \nonumber
                 &=& 0 \,\,\,\,\,\,\,\,\, \mbox{elsewhere} 
\end{eqnarray*}
and similarly for the exchange parameter $J_{\alpha \beta}$;
moreover we will neglect $U_{v v}-J_{v v}$ with respect to $U_{v v}$.
In this way the hamiltonian \ref{hhk} reduces to 
\begin{eqnarray}
\label{hhv}
\hat{H}^H  &=&  
\sum_{{\bf k} n \sigma}\epsilon^H_{{\bf k}n\sigma}
\hat{a}^{n^{\dagger}}_{{\bf k}\sigma}\hat{a}^n_{{\bf k}\sigma} 
+  \sum_{\alpha\beta} U_{\alpha \beta}
\sum_{{\bf k}{\bf k}'{\bf p}}
\sum_{n n'} \sum_{m m'}
\sum_{\sigma}
\frac{1}{2N}\cdot \nonumber \\
& &\cdot 
C_{\alpha \sigma}^n({\bf k})^{*}
C_{\alpha \sigma}^{n'}({\bf k}+{\bf p})
C_{\beta -\sigma}^m({\bf k}')^{*}
C_{\beta -\sigma}^{m'}({\bf k}'-{\bf p})
\hat{a}^{n\dagger}_{{\bf k} \sigma}
\hat{a}^{ n'}_{{\bf k}+{\bf p} \sigma}
\hat{a}^{m\dagger}_{{\bf k}'-\sigma}
\hat{a}^{m'}_{{\bf k}'-{\bf p} -\sigma}
\end{eqnarray}
Finally  we will exclude configurations with
e-h pair added to the majority-spin band as it would be strictly
correct in the strong ferromagnetic limit where no empty states are 
available in the majority-spin band. 

The states which define the basis set for the $N_e-1$ system are
schematically illustrated in fig. 1; the non-zero interaction potentials
responsible of $h-h$ and $e-h$ scattering are also indicated. Notice that
holes and electrons of parallel spin do no interact due to the assumption
$U_{v v} - J_{v v} \simeq$ 0. 

We have already stressed that 
the Hubbard correlation enters the definition of quasiparticle energies
twice: first as a mean-field correction to {\it bare } eigenvalues, 
transforming them into {\it band } ones according to eq. (\ref{eek}); second 
through the addition of self-energy (\ref{sigmab}).  
The calculation of this last quantity requires a generalization of the 
method illustrated 
in detail  in ref. \cite{Calandra} for the much simpler case of a 
single-band hamiltonian. The situation here is complicated by the sums over 
orbital indices appearing in the effective hamiltonian. This requires
the definition of orbital-dependent  diagonal Green functions and T-matrices
as described in appendix \ref{sec-app2}. 
As a result the self-energy correction to a band eigenvalue of 
wavevector ${\bf k}$ spin $\sigma$ and band index $n$ turns out 
to depend on the quantum numbers $n$ and ${\bf k}$ as  
\begin{equation}
\label{1}
\Sigma^-_{{\bf k} n\sigma}(\omega) = 
\sum_{ \beta} |C^n_{ \beta \sigma}({\bf k})|^2
\left [ \sum_{\alpha} U_{\alpha \beta} 
\frac{1}{N}\sum^{empty}_{\bf k' n'}|C^{n'}_{\alpha -\sigma}({\bf k'})|^2 -
\Sigma_{\beta \sigma}(\omega) \right ]
\end{equation}
The k-vector and band-index dependence of self-energy is associated to 
the local orbital coefficients which modulate an orbital self-energy
\begin{equation}
\Sigma^-_{ \beta \sigma}(\omega) = \sum_{\alpha}
 \int_{E_f}^{\infty} d \epsilon \;
n_{\alpha -\sigma}(\epsilon) \; T_{h-h}^{\alpha  \beta} (\omega-\epsilon) 
[1+ U_{\alpha \beta} A^{\alpha  \beta}(\omega-\epsilon)] 
\label{2}
\end{equation}
where 
$n_{\alpha \sigma} (\epsilon)$ is the spin-dependent orbital density 
of $d$  valence states and 
$T_{h-h}^{\alpha \beta}$ is the orbital dependent t-matrix describing the 
hole-hole multiple scattering
\begin{equation}
T_{h-h}^{\alpha \beta}(\omega) = \frac {U_{\alpha \beta} } {1 + U_{\alpha \beta}
 g_3^{\alpha \beta}(\omega)}
\label{3}
\end{equation}
with
\begin{equation}
g_3^{\alpha  \beta}(\omega) = 
\int_{-\infty}^{E_f} d \epsilon' \int_{-\infty}^{E_f} d \epsilon
 \frac{n_{\alpha -\sigma}(\epsilon) n_{\beta}(\epsilon')} 
{\omega -\epsilon' -\epsilon - i\delta}
\label{4}
\end{equation}
$A^{\alpha \beta}$ includes the hole-electron scattering; it is determined
by solving the integral equation
\begin{equation}
A^{\alpha \beta}(\omega,\epsilon) = 
B^{\alpha \beta}(\omega,\epsilon) + 
\int_{E_f}^{\infty} d \epsilon' \;
n_{\alpha -\sigma} (\epsilon') \;  K^{\alpha \beta}(\omega,\epsilon,\epsilon') 
A^{\alpha \beta}(\omega,\epsilon') 
\label{5}
\end{equation}
where 
\begin{equation}
K^{\alpha \beta}(\omega,\epsilon,\epsilon') = 
\int_{-\infty}^{E_f} d \epsilon'' \; n_{\alpha -\sigma}(\epsilon'')\;  
g_2^{ \beta}(\omega+\epsilon''-\epsilon) 
T_{h-e}^{\alpha \beta}(\omega+\epsilon'') 
g_2^{ \beta}(\omega+\epsilon''-\epsilon') 
T_{h-h}^{\alpha \beta}(\omega-\epsilon'')
\label{6}
\end{equation}
\begin{eqnarray}
B^{\alpha \beta}&(\omega,\epsilon)& = 
\int_{-\infty}^{E_f} d \epsilon' \; n_{\alpha -\sigma}(\epsilon')\;  
g_2^{ \beta} (\omega+\epsilon'-\epsilon) 
T_{h-e}^{\alpha \beta}(\omega+\epsilon') \times 
\nonumber\\
&& [g_1^{\alpha \beta}(\omega-\epsilon') + 
\int_{E_f}^{\infty} d \epsilon'' \;  n_{\alpha -\sigma} (\epsilon'') \; 
g_2^{ \beta}(\omega+\epsilon'-\epsilon'') g_3^{\alpha \beta}(\omega-\epsilon'') 
T_{h-h}^{\alpha \beta}(\omega-\epsilon'') ]
\label{7}
\end{eqnarray}
$T_{h-e}^{\alpha \beta}$ is the orbital dependent 
t-matrix describing the hole-electron scattering
\begin{equation}
T_{h-e}^{\alpha \beta}(\omega) = \frac {-U_{\alpha \beta}} {1 - U_{\alpha \beta}
 g_1^{\alpha  \beta}(\omega)}
\label{8}
\end{equation}
with
\begin{equation}
g_1^{\alpha  \beta}(\omega) = \int_{-\infty}^{E_f} d \epsilon' 
\int_{E_f}^{\infty} d \epsilon
 \frac{n_{\alpha -\sigma}(\epsilon) n_{ \beta \sigma}(\epsilon')} 
{\omega -\epsilon' +\epsilon - i\delta}
\label{9}
\end{equation}
and finally 
\begin{equation}
g_2^{\beta}(\omega) = \int_{-\infty}^{E_f} d \epsilon' 
 \frac{n_{\beta \sigma}(\epsilon')} {\omega -\epsilon' - i\delta}
\label{10}
\end{equation}
Equations (\ref{1} - \ref{10}) describe the procedure we have followed to 
calculate in practice self-energy corrections for the valence states of Ni 
reported in sec. \ref{sec-res}.

\section{Core states}
\label{sec-core}
The localized character of core states is responsible for much stronger
correlation effects associated with larger on-site Coulomb and exchange
integrals. In order to adapt the hamiltonian (\ref{hhk}) to the case of core
states we make the following assumptions: we neglect the
correlation among valence electrons and assume them to be described by a
single band. In (\ref{hhk}) the band index $n$ can have just two values
and as does the orbital index and we are left with a two-band hamiltonian 
\begin{equation}
\hat{H}^H  = \hat{H}_{v v} + \hat{H}_{c c} + \hat{H}_{c v}
\label{hcore}
\end{equation}
where $ \hat{H}_{v v} $ is the band hamiltonian for valence electrons
with the e-e
interaction treated in the mean-field approximation
\[
\hat{H}_{v v} = 
\sum_{{\bf k}  \sigma}\epsilon^{MF}_{{\bf k} v\sigma}
\hat{a}^{\dagger}_{{\bf k} v \sigma}\hat{a}_{{\bf k} v \sigma} 
\]
and 
\[
\hat{H}_{c c}  =  
\sum_{{\bf k}  \sigma} \epsilon^H_{c \sigma}
\hat{a}^{\dagger}_{{\bf k} c \sigma}\hat{a}_{{\bf k} c \sigma} 
+ \sum_{{\bf k}{\bf k}'{\bf p}}
\sum_{\sigma}
\frac{U_{c c}}{N}
\hat{a}^{\dagger}_{{\bf k} c \sigma}
\hat{a}_{{\bf k}+{\bf p} c \sigma}
\hat{a}^{\dagger}_{{\bf k}'c -\sigma}
\hat{a}_{{\bf k}'-{\bf p}  c -\sigma} 
\]

\begin{eqnarray*}
\hat{H}_{c v}  &=&  
\sum_{{\bf k}{\bf k}'{\bf p}}
\sum_{\sigma}\frac{1}{N} \left [U_{c v}
C_{v -\sigma}({\bf k}')^{*}
C_{v -\sigma}({\bf k}'-{\bf p})
\hat{a}^{\dagger}_{{\bf k} c \sigma}
\hat{a}_{{\bf k}+{\bf p} c \sigma}
\hat{a}^{\dagger}_{{\bf k}' v -\sigma}
\hat{a}_{{\bf k}'-{\bf p} v -\sigma}\right. \nonumber \\
&+&
\left.(U_{c v}-J_{c v})
C_{v \sigma}({\bf k}')^{*}
C_{v \sigma}({\bf k}'-{\bf p})
\hat{a}^{\dagger}_{{\bf k} c \sigma}
\hat{a}_{{\bf k}+{\bf p} c  \sigma}
\hat{a}^{\dagger}_{{\bf k}' v \sigma}
\hat{a}_{{\bf k}'-{\bf p} v \sigma}  \right ]
\end{eqnarray*}
Here
$U_{c c} = J_{c c}$, $U_{c v}$,  and
$J_{c v}$ describe the interactions between core - core  and core - valence
{\it d} electrons.

Figure 2 shows schematically the configurations to be included for the 3BS
description. Notice that configurations with one e-h pair added to the
majority-spin band are now considered: even if the number of empty states
available is small, the stronger value of the interactions  makes these
scattering channels no more negligible.  As shown in fig. 2 we have then to
take into account scattering between particles of parallel spin
proportional to $U_{c v} - J_{c v}$ and between particles of opposite spin
proportional to $U_{c v}$. The extra configurations where all the particles
(holes and electrons) have the same spin are called $|z>$. 

The procedure to calculate self-energy for core states is a straightforward
extension of the one outlined for valence states. The core hole propagator
turns out to be given by 
\begin{equation}
\label{fzz'}
{\cal G}^-\left({\bf k} c \sigma,\omega\right) =
\frac{1}{\displaystyle \omega-\epsilon^{MF}_{c \sigma} 
- \sum_{tt'}F_{3tt'}V_{t's}V_{st}
- \sum_{zz'}F_{3zz'}V_{z's}V_{sz}}
\end{equation}
The presence here of extra configurations and extra interactions does not 
imply any major  difference  with respect to the case of valence states -
just the addition of an extra term in the denominator of (\ref{fzz'})  and
the necessity to solve separately two Faddeev problems to calculate 
$F_{3 t t'}$ and $F_{3 z z'}$ for opposite and parallel spin interactions
respectively. This is a consequence of the fact that the hamiltonian
(\ref{hcore}) does not mix $|z>$ and $|t>$ configurations. The Faddeev
problem for the determination of $F_{3 z z'}$ is solved in the same way as
described in appendix \ref{sec-app2} for $F_{3 tt'}$, substituting $U_{c v}$
with $U_{c v} - J_{c v}$ and parallel spin instead of opposite ones. 
Moreover the description of the core state in terms of a zero-width band
with a delta function as orbital density of states drastically reduces the
computational effort required to evaluate the non-interacting
Green functions  (\ref{4}, \ref{9}, \ref{10}). 

The relationship between the energy $\epsilon^{MF}_{c \sigma}$ and 
the bare core eigenvalue involves as usual $Q^n_{{\bf k} \sigma}$ 
(see eq. \ref{eek} and \ref{qmf} ); explicitly one has 
\begin{equation}
\epsilon^{MF}_{c \sigma} =
\epsilon^H_{c \sigma} + 
\left[ (U_{c v} - J_{c v})<\hat{n}_{ v \sigma}> 
+ U_{c v} <\hat{n}_{v -\sigma}> \right] +  U_{c c} <\hat{n}_{c -\sigma}>
\label{ecore}
\end{equation}

It is interesting to make a comment concerning the role of the Coulomb
repulsion $U_{c c}$. Since any 3-particle (1 hole plus one e-h pair)
configuration must involve empty states, no multiple scattering
is associated  with $U_{c c}$ and this quantity gives rise just to a mean
field contribution. Core-valence interaction $U_{c v}$ and $J_{c v}$ on the
contrary modify the bare core energies both through the mean field
contribution  and by self-energy corrections which originate from hole-hole
and electron-hole scattering of opposite and parallel spin described by 
$F_{3tt'}$ and $F_{3zz'}$ respectively.

\section{Results}
\label{sec-res}
The calculation of self-energy corrections requires as an input i) the
mean-field eigenvalues and eigenvectors for valence electrons, ii)the
energy of the core level and iii) the values of the Coulomb and
exchange parameters $U_{vv}$, $U_{cc}$, $U_{cv}$, $J_{cv}$. All these
quantities can be deduced from an ab-initio LDA calculation; 
Coulomb integrals in particular can be obtained in the so-called
constrained-density functional scheme \cite{Dederichs} - a procedure which is
however not free of ambiguites (see for instance the discussion reported in
ref. \cite{Steiner}) and can lead to large variations in the estimated 
values. We have therefore adopted  a mixed strategy, using results of
ab-initio band calculations to get quantities i) only, treating all the
others as free parameters. 

The band structure of ferromagnetic nickel   has been
calculated with the linear muffin-tin orbital   (LMTO) method in the
atomic-spheres approximation (ASA)   including the combined correction term
\cite{ref1}.   The tight-binding LMTO basis set \cite{ref2} has been used,
including 9 orbitals ($s$, $p$ $d$) per atom.   The resulting energy 
dispersion is shown in fig. 3;  the occupation numbers for valence {\it d}
orbitals turns out to be $<n_{v\uparrow}> =$ 4.68, $<n_{v \downarrow}> =$ 4.07. 
We have used the single-particle eigenvalues and the corresponding
{\it d} contribution to eigenfunctions and orbital densities of states 
to calculate self-energy corrections and spectral functions
according to the theory described in the previous sections. 
A value of $U_{vv} = $ 2 eV has been chosen in order to reproduce the
observed energy position of the valence band satellite; as
we will show below we are able in this way to reproduce also other
characteristics of the valence quasi-particle states such as band width,
quasi-particle energy dispersion and exchange splitting; this is an
important result and represents a success of the present approach: previous
methods based on a semplified description of the scattering channels
\cite{Liebsch,Manghi-ni} in fact have not been able to reproduce at the
same time the satellite energy position and the valence band width which
turned out to be systematically overestimated for values of the Coulomb
integral fixed to reproduce the satellite binding energy. 

Fig. 4  shows the comparison between our results and recent
angle-integrated / spin-resolved photoemission data \cite{See}. We find
that  the calculated total spectral function 
$D^{-}_{\sigma} (\omega)$ defined as
\[
D^{-}_{\sigma} (\omega) = \sum_{{\bf k} n } D^{-}_{{\bf k} n \sigma} (\omega)
\]
closely reproduces the experimental energy distribution curves for each
spin component; notice in particular that the 6 eV satellite is observed
clearly only in the photoemission from majority-spin states in agreement
with our results. The two approximations we have adopted for the
description of valence states, that is $U_{vv}-J_{vv} \simeq 0$ and the
strong ferromagnetic limit make the self-energy corrections exactly zero
for minority-spin bands and the comparison between our results and the
spin-resolved experimental data confirms the validity of both these
assumptions. As a further evidence of this we report in fig. 4  the
minority-spin spectral function calculated after removing the assumption of
strong ferromagnetism, i.e. considering also e-h pairs added to the
majority-spin band; the small number of empty states and the relatively
small value of the interaction $U_{vv}$ make these scattering channels not
efficient and the calculated spectrum is not significantly altered. 
 
It is possible to perform a more refined analysis by looking at the k-
resolved spectral function; fig. 5 (a) shows the spectral function for the
$K$ point and $\epsilon_{{\bf k} n \uparrow}^{MF}$ = -3.6 eV together
with the corresponding self-energy $\Sigma^-_{{\bf k} n \uparrow}
(\omega)$. The spectral function shows a quasi-particle peak plus a
satellite: the first structure is associated to the pole of the hole
propagator shown in fig. 5 (b) as the interception between $Re(\Sigma^-_{{
\bf k} n \uparrow}(\omega))$ and the straight line $
\omega -\epsilon_{{\bf k} n \uparrow}^{MF} $. The second structure is
associated to the maximum of $Im(\Sigma^-_{{\bf k} n \uparrow} (\omega))$ 
and as such it will occur at the same energy at any k-vector. This
appears more clearly by extending the same analysis to the k-points along
the high symmetry line of the Brillouin Zone and plotting the energy
position of the maxima of $\displaystyle D^{-}_{{\bf k}  \sigma}(\omega) = 
\sum_ n D^{-}_{{\bf k} n \sigma} (\omega) $ to get the
quasi-particle band structure of fig. 6. By comparing the quasi-particle
band structure with the single-particle results it appears that the
majority-spin eigenvalues are heavily affected by self-energy corrections,
showing a strong reduction of the {\it d} - band width and the presence of
the above mentioned 6 eV satellite. Since the majority-spin eigenvalues are
shifted to lower binding energies while the minority-spin ones are
unaffected, the splitting between majority and minority states becomes
smaller than in the original single particle bands. This goes in
the right direction since it is well known
that single particle calculations overestimate this quantity; 
from fig. 6 it appears that the energy separation
between majority and minority quasiparticle peacks around $\Gamma$ point is
0.2 eV for the topmost band. This result is in remarkable agreement with a
recent estimate reporting a split of 204$\pm$8 meV along the $\Sigma$
direction \cite{Kreutz}. The same is true for the quasi-particle energy
dispersion as a whole, which is shown in fig. 7 compared with the results of
angle-resolved spin-integrated photoemission spectroscopy of ref.
\cite{sakisaka}. 
 
As far as core levels are concerned we consider here the {\it 3p} level of
nickel as a test case. Since the core levels of an isolated atom are 
degenerate in spin their spin dependence is a solid state effect,
associated to the interaction between the atom and the solid it is
embedded in; in other words the spin dependence of core level energies is 
related to the spin polarization of the valence band on one side and to
core-valence interactions on the other. According to our view the relevant
quantities are the Coulomb and exchange integrals between  core and valence
orbitals which affect the bare atomic core energies both through the mean
field term $Q^n_{{\bf k} \sigma}$ and self-energy corrections. The
mean-field contribution of eq. (\ref{ecore}) gives rise to a spin splitting
proportional to $J_{cv}$ and to the valence band spin polarization 
\[
\epsilon^{MF}_{c \uparrow} - \epsilon^{MF}_{c \downarrow} =
J_{c v}\left (<\hat{n}_{ v \downarrow}> - <\hat{n}_{ v \uparrow}> \right )
\]
Notice that $U_{c v}$ and $U_{cc}$ do not enter this expression; they
affect the bare core level $\epsilon^H_{c \uparrow} = \epsilon^H_{c
\downarrow} $ according to 
\begin{equation}
\epsilon^H_{c \uparrow} + 
\left[ U_{c v} \left (<\hat{n}_{ v \uparrow}> +<\hat{n}_{ v \downarrow}> 
\right ) + 
 U_{c c} <\hat{n}_{c \downarrow}> \right] \equiv \epsilon_c
\end{equation}
giving rise to a modified core energy level $\epsilon_c$ which is still
spin-independent. The mean-field eigenvalues are related to this quantity 
as 
\[
\epsilon^{MF}_{c \sigma} = \epsilon_c - J_{c v} <\hat{n}_{ v \sigma}> 
\]
We have used spin-orbit split values of $\epsilon_c$, 
that is $\epsilon_c^{3/2}$ and $\epsilon_c^{1/2}$,  to
fix the absolute value of the core level binding energy, choosing then
$J_{cv}$ to reproduce the spin-splitting of the main peak observed in the
core-level photoemssion \cite{See}. The last parameter, that is $U_{cv}$
has been fixed in order to reproduce the satellites energy position. The
values which optimize the agreement between our calculation and the
experimental results are listed in Table \ref{table1} 
 
Fig. 8 shows the calculated self-energy for the creation of both majority
and minority-spin hole in the core level $3p^{1/2}$. The same analysis
obviously applies to the other spin-orbit split level $3p^{3/2}$. In the
case of the  minority-spin core hole the self-energy presents two well
defined structures: the one at lower binding energy is related to the
scattering between particles of parallel spin with strength proportional to
$U_{cv}-J_{cv}$ (see fig. 2(f)) while the structure at higher binding
energy is related to $U_{cv}$ and to the scattering between opposite spin
particles (see fig. 2 (e)). Notice that two independent factors determine
the efficiency of the scattering process, the strength of the interaction
on one side and the number of available states in the valence band on the
other. In the case of a minority-spin core hole the weaker parallel spin
interaction is compensated by the larger number of empty valence states and
both the scattering channels involving parallel and opposite spin particles
play a role. The same argument applies to the case of majority-spin core
hole but now the weaker parallel spin interaction is associated to a
small number of available empty valence states (see fig. 2 c); for this
reason the interaction between opposite spin particles (see fig. 2 (b))
remains the only efficient scattering channel. As a consequence the
self-energy for majority-spin core hole presents a single structure
associated to $U_{cv}$. 

As discussed in the previous section satellite structures are expected to
occur at energies where the imaginary part of self-energy has a maximum
giving rise to a complex pole of the hole propagator and therefore to a
short lived excitation. The structures in the
calculated self-energy 
we have just described and their origin are therefore essential in order to
get a physical interpretation of the observed photoemission spectrum. We
report in fig. 9 the calculated spin-integrated spectral density for the
creation of a {\it 3p} core hole compared with the photoemission results of
ref. \cite{Liu}. The spectrum shows a main peak at about 66 eV with two
characteristic spin-orbit split structures (A, B) and two satellites (C, 
D). The decomposition of the spectral function into contributions from the
two spin-orbit split levels $3p^{1/2}$ and $3p^{3/2}$ and from different
spins is also shown. The spectral function for the creation of a
majority-spin core hole shows a main peack and a satellite for each
spin-orbit split level; as discussed above in terms of the self-energy
this satellite is associated to the only efficient scattering channel which
comes into play after the removal of one majority- spin electron, that is
to configurations of fig. 2. (b) where the majority-spin core hole
interacts with opposite spin particles in the valence band, with a strength
proportional to $U_{cv}$. The spectral function for the creation of a
minority-spin core hole presents instead two satellites for each spin-orbit
split level: the one at the higher binding energy is associated to the
configurations of fig. 2 (e) where the spin-down core hole interacts with
opposite spin particles in the valence band with a strength proportianal to
$U_{cv}$. Due to the small number of empty states available in the spin-up
band this satellites is less pronounced here than in the majority-spin
spectrum. The satellite at lower binding energy is related to
configurations of fig. 2 (f) and to scattering between parallel spin
particles of strenght proportional to $U_{cv}-J_{cv}$. 

The one-to-one correspondence between the configurations of fig. 2 and the
satellite structures allows us to interpret them as as shake-up processes
occurring after the removal of either a minority or a majority-spin core
electron. The satellite at lower binding energy (C) is then associated to the
creation of one minority-spin  core-hole plus  an e-h pair in the valence
band of the same spin, giving rise to a bound state of three parallel spin
particles  which can be defined AS a triplet state; the satellite at higher
binding energy (D) is related to the creation of either a majority or a
minority-spin core hole plus a valence  e-h pair of opposite spin giving
rise to a singlet state. 

To analyze the spin polarization of the whole core hole spectrum in more
details 
it is useful to consider the spin-resolved spectra and their 
difference $D^-_{\uparrow}(\omega) - D^-_{\downarrow}(\omega)$
reported in fig. 10. 
It appears that the spin polarity
of structures A, B, C and D of fig. 9 is "down", "up", "down", "up"
respectively. This is again in agreement with what has been seen 
experimentally and reported in ref. \cite{Carbone,Liu,Saitoh}.

\section{Summary}
\label{sec-sum}
We have described a method for including short-range on-site interactions
in the description of both valence and core states of a solid system. When
applied to valence states of ferromagnetic nickel the method allows to get
a quasiparticle band structure which compares much more favourably with the
experimental observation than conventional mean-field LDA band structure
eigenvalues , reproducing the observed band width, the energy dispersion,
the satellite structure and the exchange splitting. Since the method does
not rely on a perturbation expansion it has a wide range of application,
including any correlation regime. The extension to core levels is quite
natural and allows to take fully into account the itinerant character of
valence electrons. We get then a physical picture of the {\it 3p} core level
spectrum where the spin splitting of the main line is associated to the
valence band spin polarization and to the core-valence exchange
interaction; the two satellites are interpreted as arising from
shake-up processes
occurring after the removal of a core electron, involving the creation of a
{\it d} band e-h pair. 
Even though our present choice of empirically determining the parameters of
the Hubbard hamiltonian ensures that we obtain an overall good agreement
with experiments we believe that the possibility of reproducing both the
satellite structures, the main line and their spin dependence with just
four parameters can be seen as a non trivial result. The widely used
atomic models - which use empirical parameters as well -
require also an ad-hoc evaluation of the so called extra-atomic effects
which allow to take into account the role of electrons on neighbouring
atoms \cite{Seeprl}; such effects are on the contrary included here from
the beginning since the full details of the valence band structure of the
solid system are considered. In this sense the present approach can be
considered an appropriate tool to describe the response of an itinerant
electron system to the creation of a core hole. 

\appendix
\section{Matrix elements of the multi-orbital Hubbard hamiltonian}
\label{sec-app1}

As discussed in section \ref{sec-val} the application of the 3BS method to
the valence band states of nickel requires the explicit definition of 
matrix elements of the Hamiltonian \ref{hhv} containing a non-interacting 
part that we call here $\hat{H}^0$
and the on-site interaction among opposite spin electrons that we denote 
by $\hat{H}'$. The matrix elements involve states $|s>$ and $|t>$ 
defined in \ref{states} and pictorially depicted in fig. 1. 
We have  
\begin{eqnarray*}
	\langle s|\hat{H}^0|s\rangle &=&   
	\sum_{\bbox{k}' n' \sigma'}^{occ}\epsilon_{
	\bbox{k}' n'
	\sigma'}^H-\epsilon_{\bbox{k} n \sigma}^H \\
	\langle s|\hat{H}^0|t\rangle &=& 0 \\
	\langle t|\hat{H}^0|t\rangle &=&
	\sum_{\bbox{k}' n' \sigma'}^{occ}\epsilon_{
	\bbox{k}' n'
        \sigma'}^H+\epsilon_{\bbox{q}_3 n_3 -\sigma}^H -
	\epsilon_{\bbox{q}_2 n_2 -\sigma}^H -
	\epsilon_{\bbox{q}_1 n_1 \sigma}^H 
\end{eqnarray*}

\begin{eqnarray*}
	\langle s|\hat{H}'|s\rangle &=&
	\sum_{\alpha\beta}\frac{U_{\alpha\beta}}{N}\sum_{\bbox{k}'' n''}
	\left |C_{\alpha -\sigma}^{n''}(\bbox{k}'') 
	\right |^2 \left \{ \sum_{
	\bbox{k}' n'}\left |
        C_{\beta\sigma}^{n'}(\bbox{k}') \right |^2-\left |
        C_{\beta\sigma}^n(\bbox{k}) \right|^2 \right \} \\
	\langle s|\hat{H}'|t\rangle &=&
        	-\sum_{\alpha\beta}\frac{U_{\alpha\beta}}{N}C_{\alpha -\sigma}^{n_2}
(\bbox{q}_2)^{*}C_{\alpha -\sigma}^{n_3}(\bbox{q}_3)C_{\beta
	\sigma}^{n_1}(\bbox{q}_1)^{*}
        C_{\beta\sigma}^{n}(\bbox{k})\delta_{\bbox{k}-\bbox{q}_1,
	\bbox{q}_2-\bbox{q}_3} \\
	\langle t|\hat{H}'|t\rangle &=&
	\sum_{\alpha\beta}\frac{U_{\alpha\beta}}{N} \left \{
        \sum_{\bbox{k}' n'}\left |C_{\alpha -\sigma}^{n'}
	(\bbox{k}') \right |^2\sum_{\bbox{k}'' n''}
        \left |C_{\beta\sigma}^{n''}(\bbox{k}'') 
	\right |^2+\left |C_{\alpha -\sigma}^{n_3}(\bbox{q}_3)\right |^2
	\sum_{\bbox{k}' n'}\left |C_{\beta\sigma}^{n'
	}(\bbox{k}')\right |^2 \right.  + \\
        & &\left. -\left |C_{\alpha -\sigma}^{n_2}(\bbox{q}_2) \right |^2
        \sum_{\bbox{k}' n'}\left |C_{\beta\sigma}^{n'
	}(\bbox{k}') \right |^2
	-\left |C_{\beta\sigma}^{n_1}(\bbox{q}_1) \right |^2
        \sum_{\bbox{k}' n'}\left |C_{\alpha -\sigma}^{n'}(\bbox{k}') 
	\right |^2\right\} \\
	\langle t'|\hat{H}'|t\rangle &=&
        -\sum_{\alpha\beta}\frac{U_{\alpha\beta}}{N}C_{\alpha -\sigma}^{
	n_3'}(\bbox{q}_3')^{*}C_{\alpha -\sigma}^{n_3}(
	\bbox{q}_3)C_{\beta\sigma}^{n_1}(\bbox{q}_1)^{*}
        C_{\beta\sigma}^{n_1'}(\bbox{q}_1')\delta_{\bbox{q}_2,
	\bbox{q}_2'}
        \delta_{n_2,n_2'}\delta_{\bbox{q}_1-\bbox{q}_3,\bbox{q}_1'
	-\bbox{q}_3'}+ \\
        & & +\sum_{\alpha\beta}\frac{U_{\alpha\beta}}{N}C_{\alpha -\sigma}^{
	n_2}(\bbox{q}_2)^{*}
        C_{\alpha -\sigma}^{n_2'}(\bbox{q}_2')C_{\beta
	\sigma}^{n_1}(\bbox{q}_1)^{*}C_{\beta\sigma}^{n_1'}(
	\bbox{q}_1')\delta_{\bbox{q}_3,\bbox{q}_3'}
        \delta_{n_3,n_3'}\delta_{\bbox{q}_1+\bbox{q}_2,
	\bbox{q}_1'+\bbox{q}_2'}	
\end{eqnarray*}

As discussed in section \ref{sec-core} 
the description of core states requires the inclusion of the interaction 
involving the exchange integral that we call here $\hat{H}''$
and the extension of the complete set to the three-particle 
configurations $|z>$ of fig. 2. It is easy to show that in this case the 
diagonal matrix elements are similar to the previous ones and that the non-
zero off-diagonal ones are 
$\langle s|\hat{H}'|t\rangle$ ,
$\langle t|\hat{H}'|t'\rangle$ ,
$\langle z|\hat{H}''|z'\rangle$ .

\section{Faddeev equations for the multi-band Hubbard hamiltonian}
\label{sec-app2}
We illustrate the procedure which leads to eq. (\ref{1}) -
(\ref{9}) for the removal of 
one spin up electron. 
The relationship (\ref{ftt'}) 
between the diagonal hole-propagator and the 3-body 
resolvent is obtained by inserting 
the completness relation
\[
\sum_s |s><s| + \sum_t |t><t|
\]
into the identity 
\[
<s|(z-\hat{H^H}) \frac{1}{z-\hat{H^H}}|s> = 1
\]
that is
\begin{equation}
[z-(H_1)_{ss}] G_{ss}(z) - \sum_t V_{s t} G_{t s}(z) = 1
\label{gts}
\end{equation}
Since $G_{t s}(z)$ can be obtained by eq. (\ref{dyson}) as
\begin{equation}
 G_{ts}(z) = \sum_{ts'} F_{tt'}V_{t's'}G_{s s'}(z)
\label{gss'}
\end{equation}
one gets
\begin{equation}
[z-(H_1)_{ss}]G_{ss}(z) - \sum_{t t'} V_{st} V_{t's} F_{tt'} G_{ss} +
\sum_{t t'} \sum_{s \neq s'} V_{st} V_{t's'} F_{tt'} G_{ss'} = 1
\label{a3}
\end{equation}
By substituting the explicit definition of the matrix elements of 
$\hat{V}$ it appears that the two summations in (\ref{a3})
involve terms of the kind
\[
\sum_{\beta  \delta}
C^{n}_{\beta \uparrow}({\bf k})
C^{n'}_{\delta \uparrow}({\bf k})^{*}
\]
with $n=n'$, $n\neq n'$ in the first and  second sum respectively. For this 
reason it seams reasonable to neglect in (\ref{a3})
the last summation with respect to 
the  first one getting in this way eq. (\ref{ftt'}) as a result. 

According to eq. (\ref{sigmab}) the self-energy is given by the sum
\begin{eqnarray}
 \sum_{tt'}F_{3tt'}V_{t's}V_{st} 
&=&
\frac{U^2}{N^2} 
\sum_{q_1  q_2 q_3}
\sum_{n_1  n_2 n_3}
\sum_{\alpha \beta }
C^{n_1}_{\alpha\uparrow}({\bf q_1})^{*}
C^{n_3}_{\alpha\downarrow}({\bf q_3})
C^{n_2}_{\beta\downarrow}({\bf q_2})^{*}
C^{n}_{\beta\uparrow}({\bf k})
\\ \nonumber
&\cdot &
\sum_{q_1'  q_2' q_3'}
\sum_{n_1'  n_2' n_3'}
\sum_{\gamma \delta} 
C^{n_1'}_{\gamma\uparrow}({\bf q_1'})
C^{n_3'}_{\gamma\downarrow}({\bf q_3'})^{*}
C^{n_2'}_{\delta\downarrow}({\bf q_2'})
C^{n}_{\delta\uparrow}({\bf k})^{*} 
\\ \nonumber
&\cdot &
<q_1 n_1\uparrow  q_2 n_2\downarrow q_3 n_3\downarrow|
|\hat{F}_3^D + \hat{F}_3^D (\hat{S}_{h-e} + \hat{T}_{h-h} \hat{F}_3^D 
\hat{S}_{h-e})\hat{F}_3^D \\ \nonumber
&\cdot &
|q_1' n_1' \uparrow  q_2' n_2' \downarrow q_3' n_3' \downarrow>
\label{a5}
\end{eqnarray}
where we have used the definition \ref{states} for $|t>$ and $|t'>$. 
Let us define now the orbital dependent free propagator
describing h-h scatterings
\[
g_3^{\alpha  \beta }({\bf q_3} n_3  \omega) =
\frac{1}{N^2} \sum_{q_1 n_1  q_2 n_2} 
\frac{ 
C^{n_1}_{\alpha\uparrow}({\bf q_1})^{*}
C^{n_1}_{\alpha\uparrow}({\bf q_1})
C^{n_2}_{\beta\downarrow}({\bf q_2})^{*}
C^{n_2}_{\beta\downarrow}({\bf q_2})}
{\omega-(E_0-
\epsilon^{MF}_{{\bf q_1}n_1 \uparrow}-
\epsilon^{MF}_{{\bf q_2}n_2 \downarrow}+
\epsilon^{MF}_{{\bf q_3}n_3 \downarrow})}
\]
where the summation is over filled states of band indeces $n_1$ and $n_2$.
By using the definition (\ref{tmat}) it is easy to show that
\begin{eqnarray*}
 &&\sum_{q_1 n_1  q_2 n_2} 
C^{n_1}_{\alpha\uparrow}({\bf q_1})^{*}
C^{n_2}_{\beta\downarrow}({\bf q_2})^{*}
<q_1 n_1\uparrow  q_2 n_2\downarrow q_3 n_3\downarrow|
\hat{G}_3^D \hat{T}_{h-h}
|q_1' n_1' \uparrow  q_2' n_2' \downarrow q_3' n_3' \downarrow> =
\\ \nonumber
&&= \frac{C^{n_1}_{\alpha\uparrow}({\bf q_1})^{*}
C^{n_2}_{\beta\downarrow}({\bf q_2})^{*}
g_3^{\alpha  \beta }({\bf q_3} \omega)}
{1-U_{\alpha \beta} g_3^{\alpha  \beta }({\bf q_3} \omega)}
\delta_{ {\bf q_3}, {\bf q_3'}}
\end{eqnarray*}
It is also useful
to define the orbital-dependent t-matrix for h-h scatterings
\[
T_{h-h}^{\alpha \beta}({\bf q_3} n_3 \omega) = \frac{U_{\alpha \beta}}{1-
U_{\alpha \beta} g_3^{\alpha \beta}({\bf q_3} n_3 \omega)}
\]
Similar definitions and relations hold for e-h scatterings.
It is then a matter of simple algebra to transform eq. (\ref{a5})
into the form
\begin{eqnarray*}
 \sum_{tt'}F_{3tt'}V_{t's}V_{st} &=&
\sum_{\beta} |C^{n}_{\beta}({\bf k}\uparrow)|^2
\\ \nonumber
&-&
\sum_{\alpha} \frac{U_{\alpha \beta}}{N} \sum_{{\bf q_3} n_3} 
[ |C^{n}_{\beta}({\bf q_3} \downarrow)|^2 
\\ \nonumber
&+& 
T_{h-h}^{\alpha \beta}({\bf q_3} n_3)
\left( 
 |C^{n}_{\beta}({\bf q_3} \downarrow)|^2 + U_{\alpha \beta} 
A^{\alpha \beta} ({\bf q_3} n_3) \right)
\end{eqnarray*}
where
\begin{eqnarray*}
A^{\alpha \beta} ({\bf q_3} n_3) &=&
\frac{1}{N}
\sum_{q_1  q_2 }
\sum_{n_1  n_2 }
C^{n_1}_{\alpha\uparrow}({\bf q_1})^{*}
C^{n_2}_{\beta\downarrow}({\bf q_2})^{*}
C^{n_3}_{\alpha\downarrow}({\bf q_3})
\\ \nonumber
& . &
\sum_{q_1'  q_2' q_3'}
\sum_{n_1'  n_2' n_3'}
C^{n_1'}_{\gamma\uparrow}({\bf q_1'})
C^{n_3'}_{\gamma\downarrow}({\bf q_3'})^{*}
C^{n_2'}_{\delta\downarrow}({\bf q_2'})
\\ \nonumber
& . &
<q_1 n_1\uparrow  q_2 n_2\downarrow q_3 n_3\downarrow|
\hat{F}_3^D \hat{S}_{h-e} \hat{F}_3^D
|q_1' n_1' \uparrow  q_2' n_2' \downarrow q_3' n_3' \downarrow>
\end{eqnarray*}
From now on the procedure is the same as the one described in ref.
\cite{Calandra}, adopting in particular the 
so-called local approximation \cite{treglia} which 
allows to transform all the summation over k-vectors into integrals
involving the density of states.

\begin{figure}
\caption{ Schematic representation of the basis set for the configuration
expansion of the interacting state with one electron removed from the
majority-spin band. $\hat{V}$, $\hat{V}_{h-h}$, $\hat{V}_{h-e}$ describe the
coupling between 1- and 3-particle states, the hole-hole and the hole-
electron interaction respectively.
\label{f1}}
\end{figure}

\begin{figure}
\caption{ Schematic representation of the basis set for the configuration
expansion of core states. 
\label{f2}}
\end{figure}

\begin{figure}
\caption{ Single particle band structure of nickel obtained using the LMTO
method. Energies are referred to the Fermi energy. Open  triangles,
minority-spin states; filled triangles, majority-spin states. 
\label{fig3}}
\end{figure}

\begin{figure}
\caption{ Density of quasi-particle states of nickel for majority-spin  (a)
and minority-spin (b) spin compared with experimental results of 
angle-integrated spin-resolved photoemission results (filled
triangles) of ref. \protect\cite{See}.
 The results for minority-spin bands obtained without the
strong-ferromagnetic-limit approximation  is shown as a broken line in
panel (b). 
\label{f4}}
\end{figure}

\begin{figure}
\caption{ Spectral function (a) and self-energy (b) for the $K$ point and
$\epsilon_{{\bf k} n \uparrow}^{MF}$ = -3.6 eV. The interception between
$Re\left(\Sigma^-({\bf k} n \uparrow,\omega) \right ) $ and the straight
line $ \omega -\epsilon_{{\bf k} n \uparrow}^{MF} $ indicate the position
of the quasi-particle pole. 
\label{f5}}
\end{figure}

\begin{figure}
\caption{ Quasi-particle band structure of nickel along the high symmetry
directions of the Brillouin Zone. Energies are referred to the Fermi
energy. Open  triangles, minority-spin states; filled triangles,
majority-spin states. 
\label{f6}}
\end{figure}

\begin{figure}
\caption{ Comparison between the calculated dispersion of hole
quasi-particle states (circles) for majority-spin bands and angle-resolved
spin-integrated photoemission results (diamonds) of reference
\protect\cite{sakisaka}.
\label{f7}}
\end{figure}

\begin{figure}
\caption{ Self-energy for the creation of majority-spin (a) and
minority-spin hole (b) in the core level $3p^{1/2}$. 
\label{f8}}
\end{figure}

\begin{figure}
\caption{ Spin-integrated spectral function for the creation of a {\it 3p}
core hole compared with the photoemission results of ref. 
\protect\cite{See} (a)
and its decomposition into contributions from the spin-orbit split level
$3p^{1/2}$ (continuous line) and $3p^{3/2}$ (dashed line). Panel (b) refers
to majority-spin state and panel (c) to the minority-spin one. 
\label{f9}}
\end{figure}

\begin{figure}
\caption{ (a) Calculated spectrum for  the creation of a majority-spin
(continuous line) and minority-spin (dashed line) {\it 3p} core hole. (b)
Difference spectrum between the two spin component. 
\label{f10}}
\end{figure}

\mediumtext
\begin{table}
\caption{Parameters used in the calculations of Ni {\it 3p} core hole 
spectrum}
\begin{tabular}{cccc}
$U_{cv}$ & $J_{cv}$ & $\epsilon_c^{1/2}$ & $\epsilon_c^{3/2}$  \\
\tableline
$5.00$ \,eV & $2.50$ \,eV & $-59.00$ \,eV & $-57.50$ \,eV 
\end{tabular}
\label{table1}
\end{table}
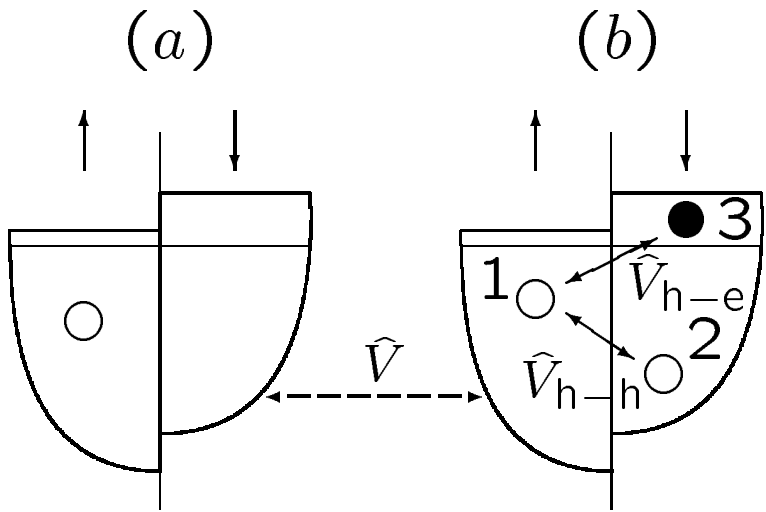
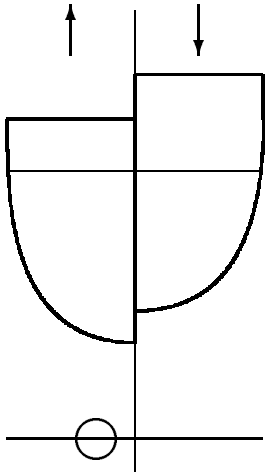
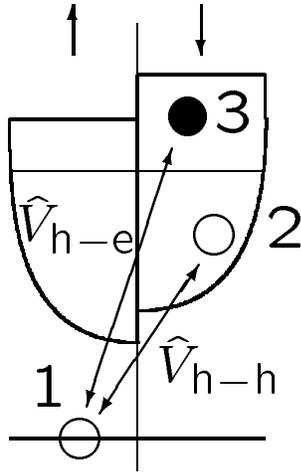
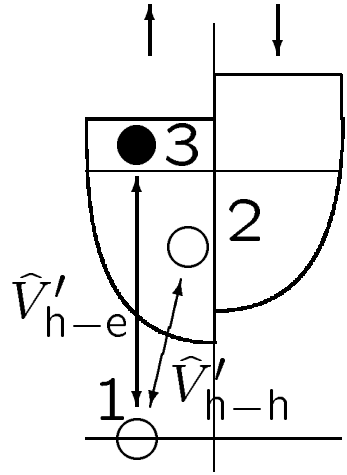
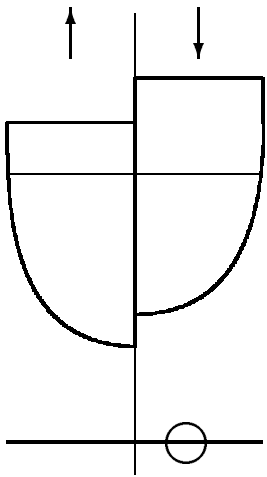
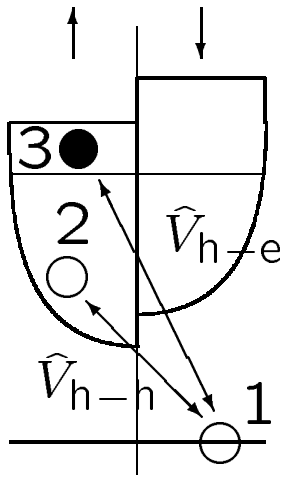
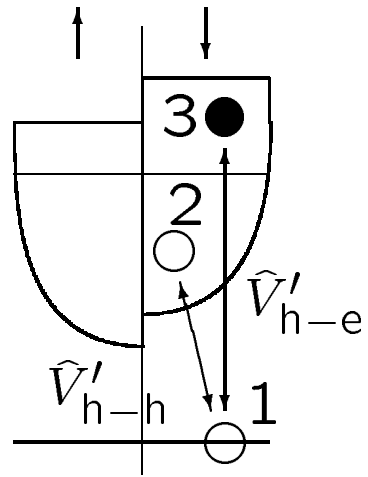
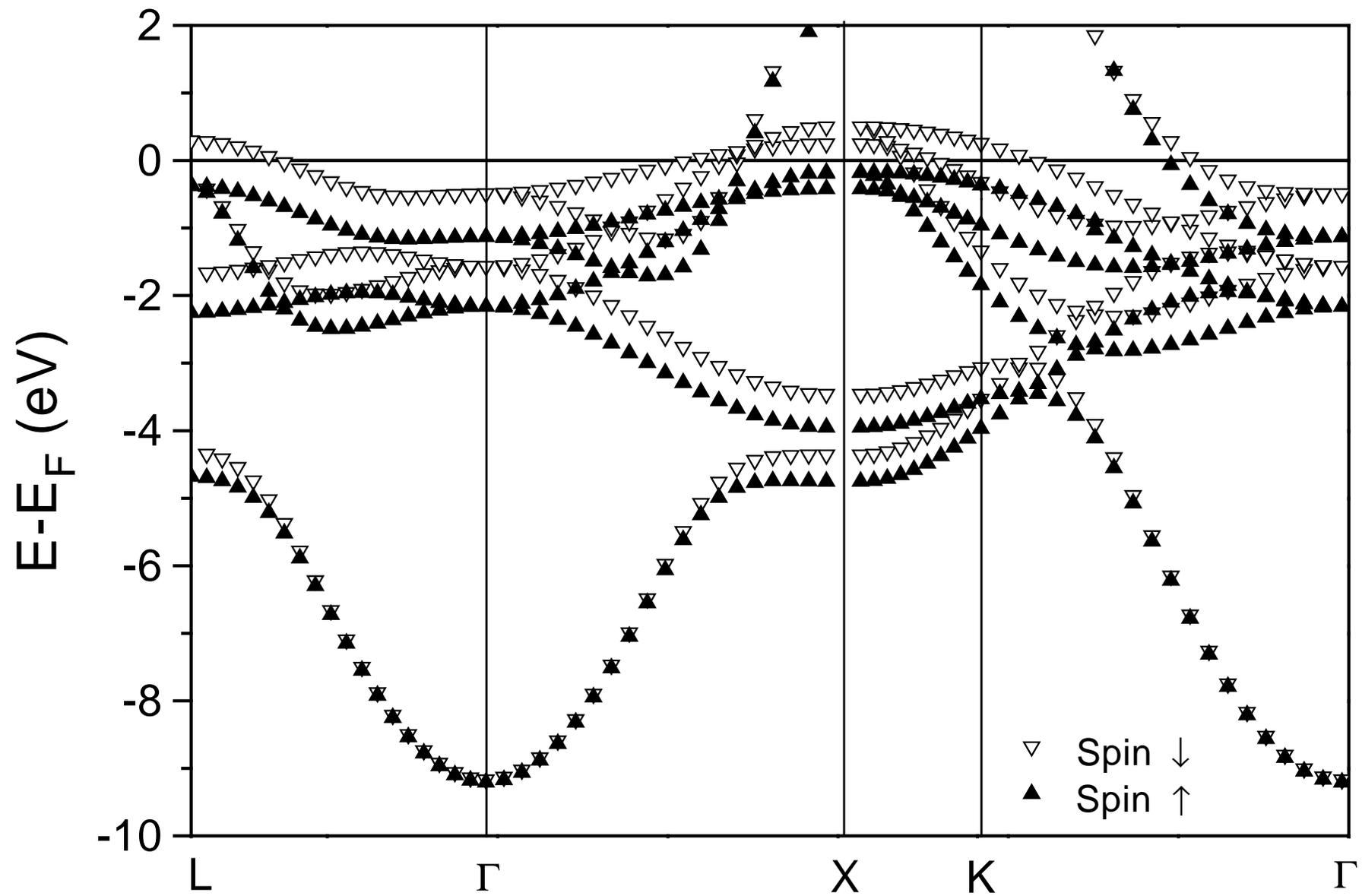
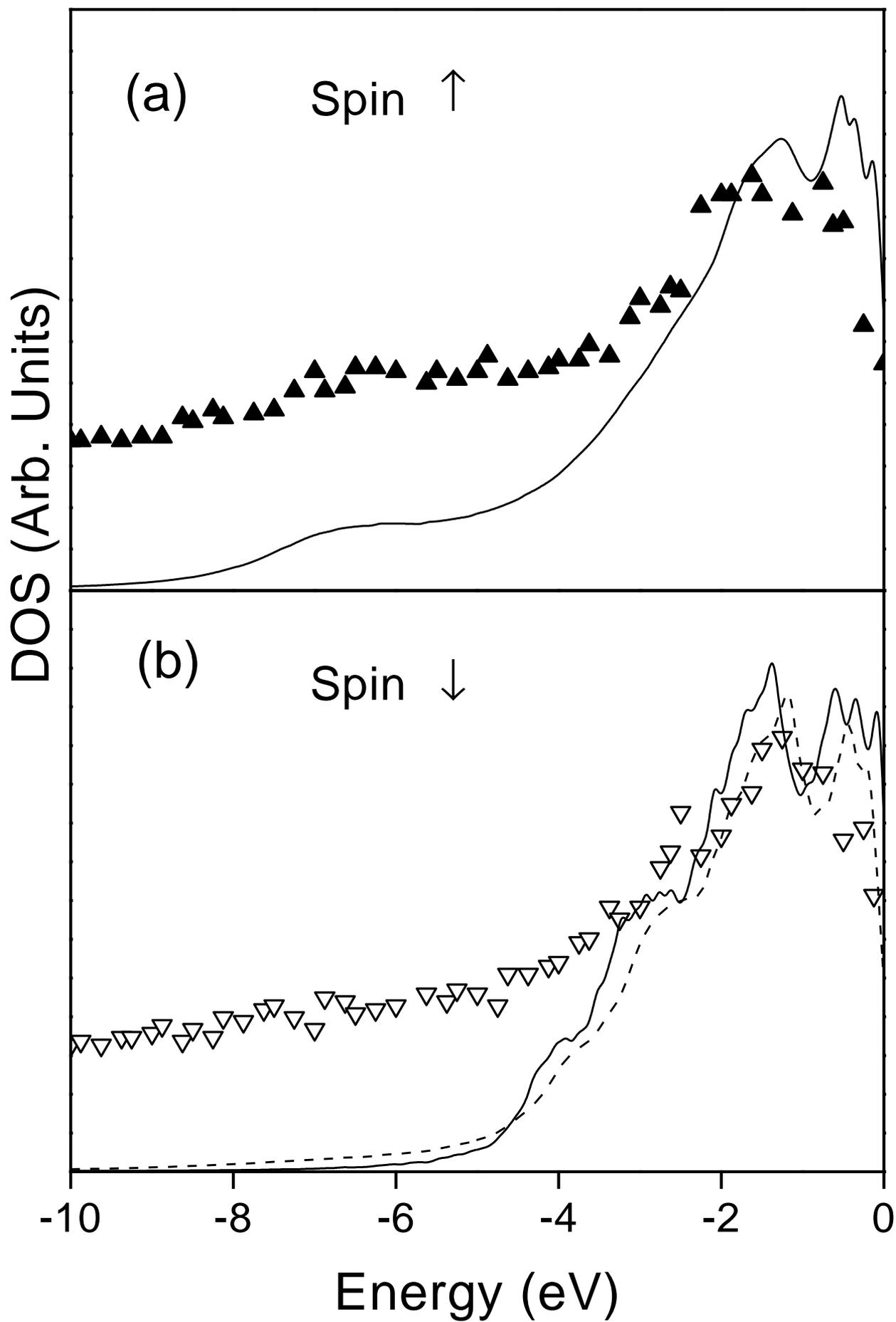
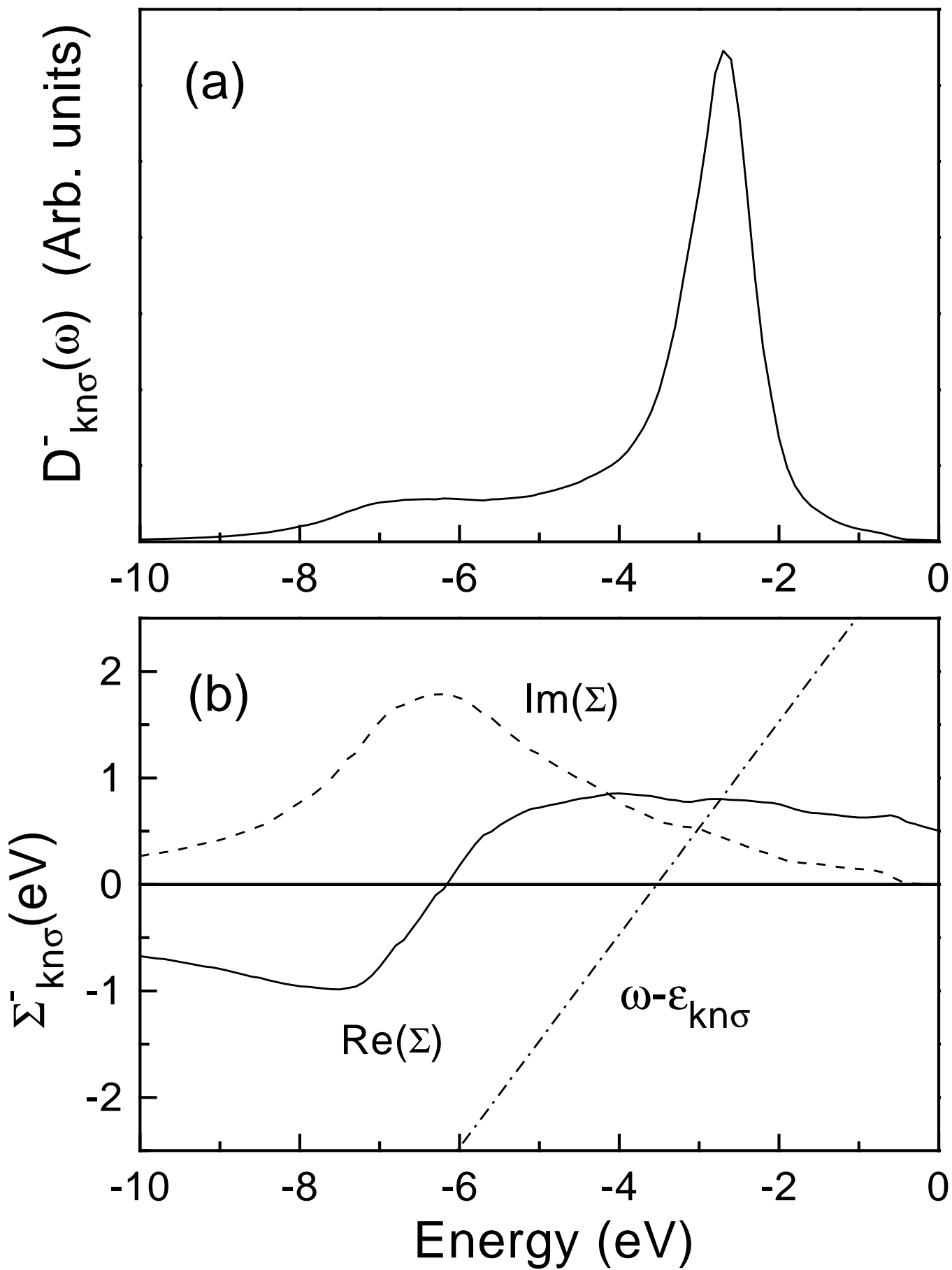
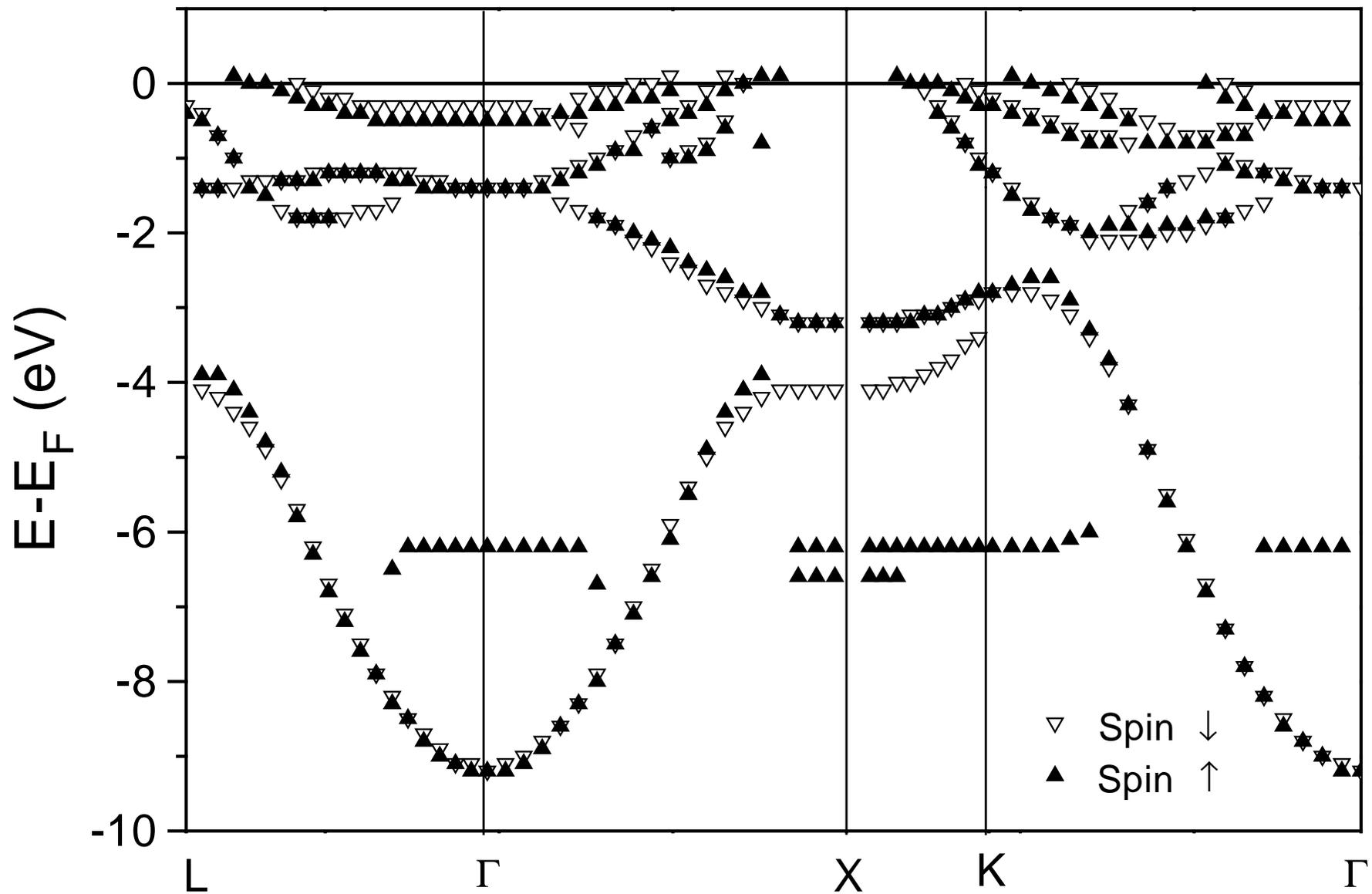
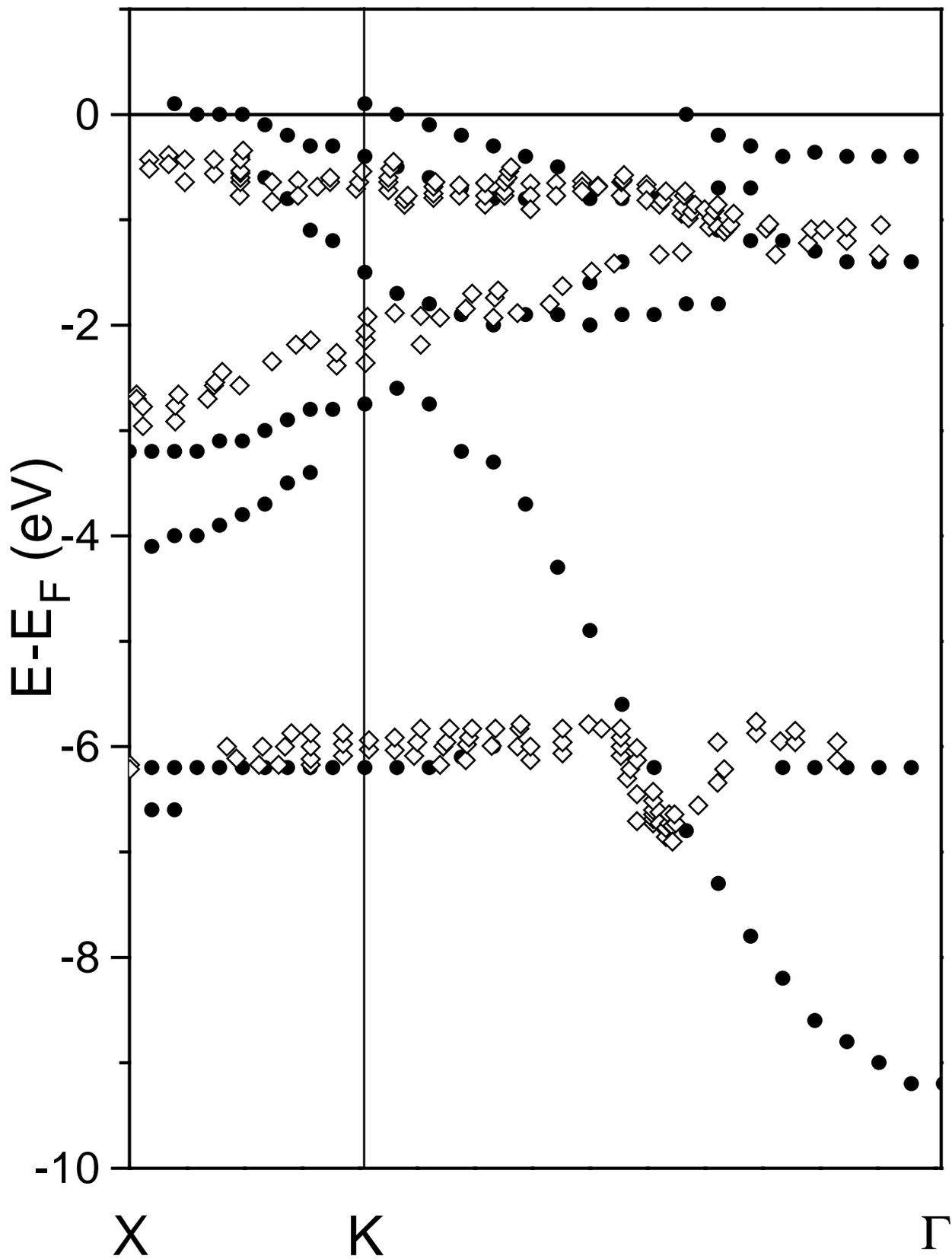
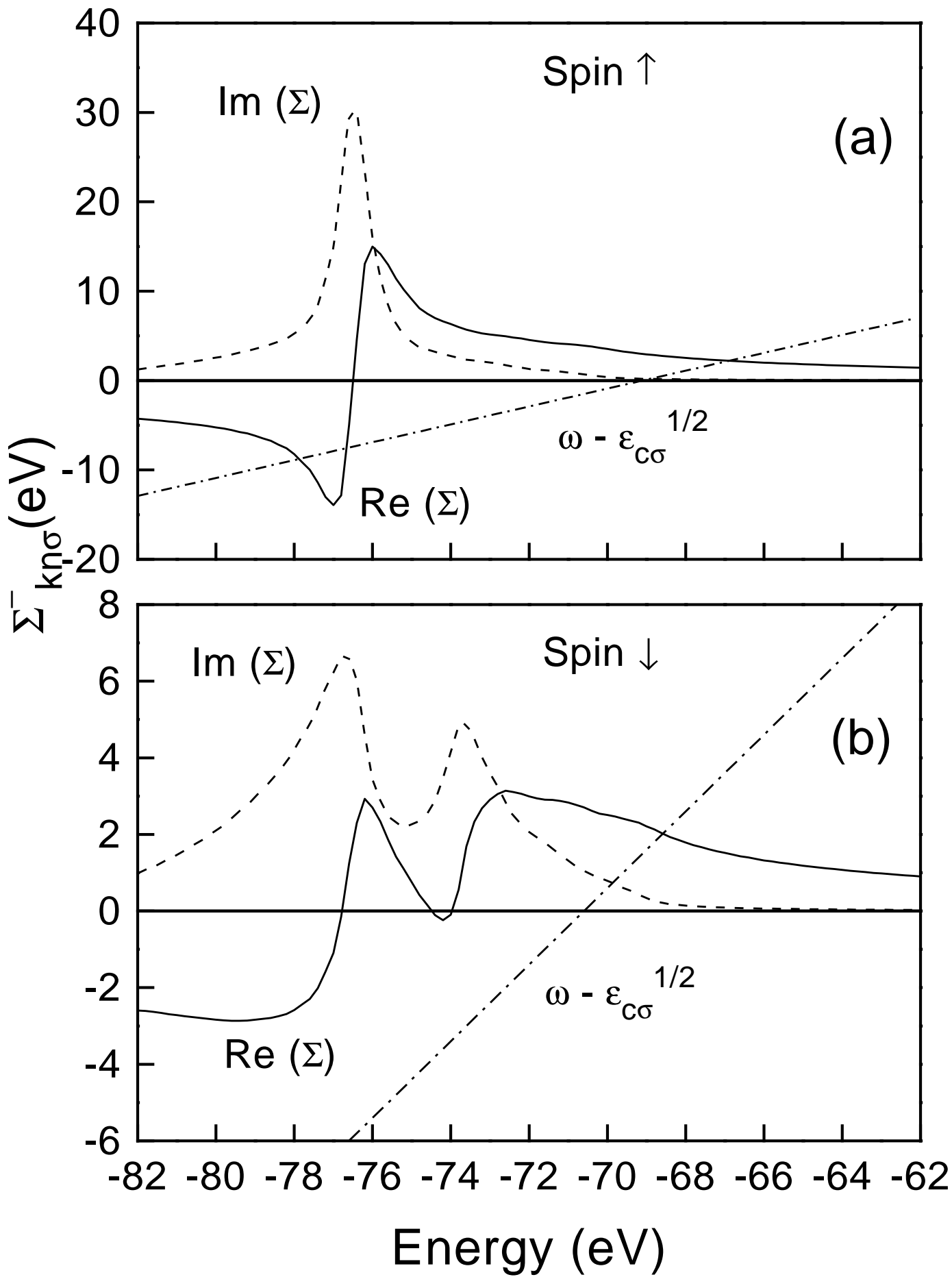
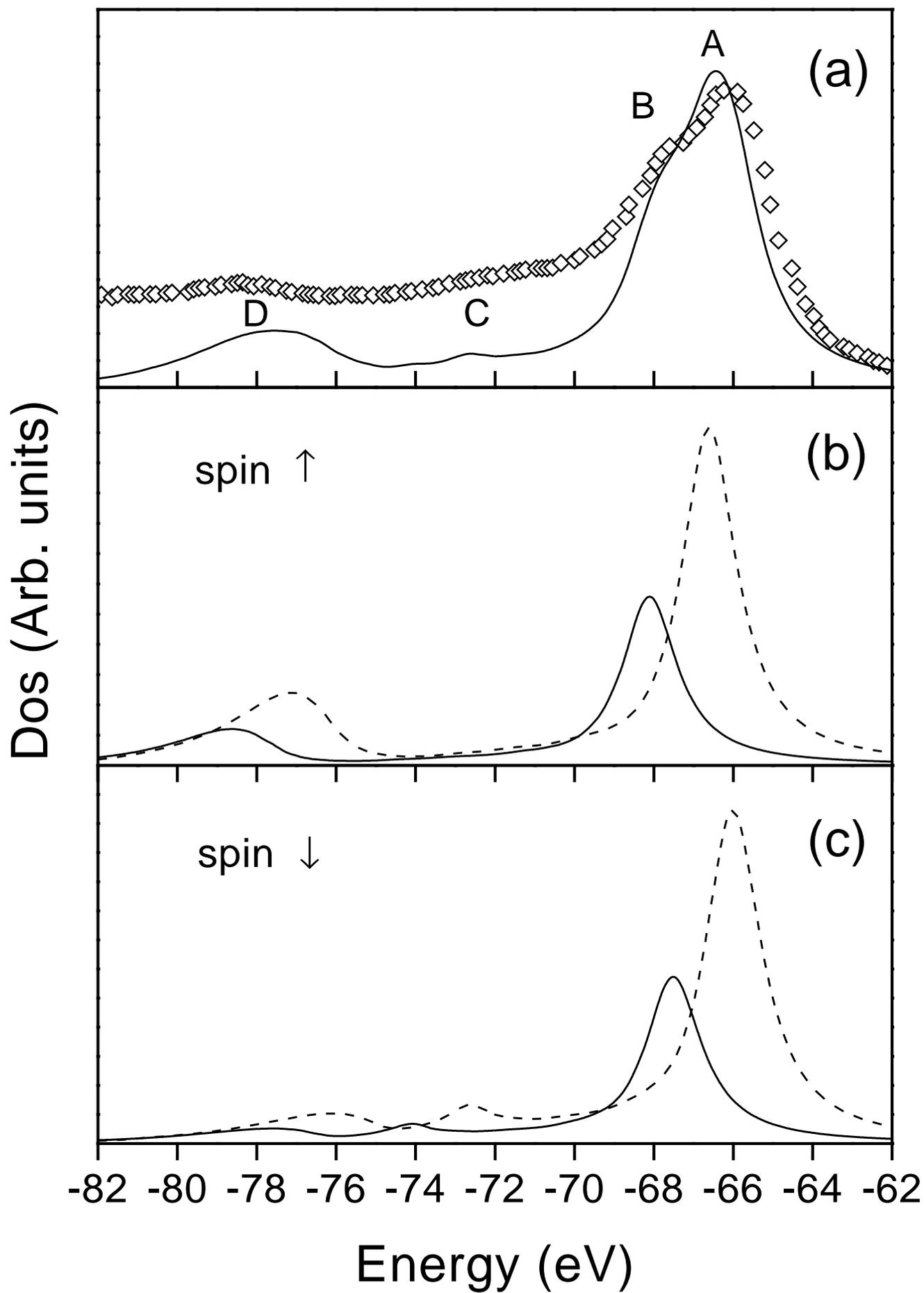
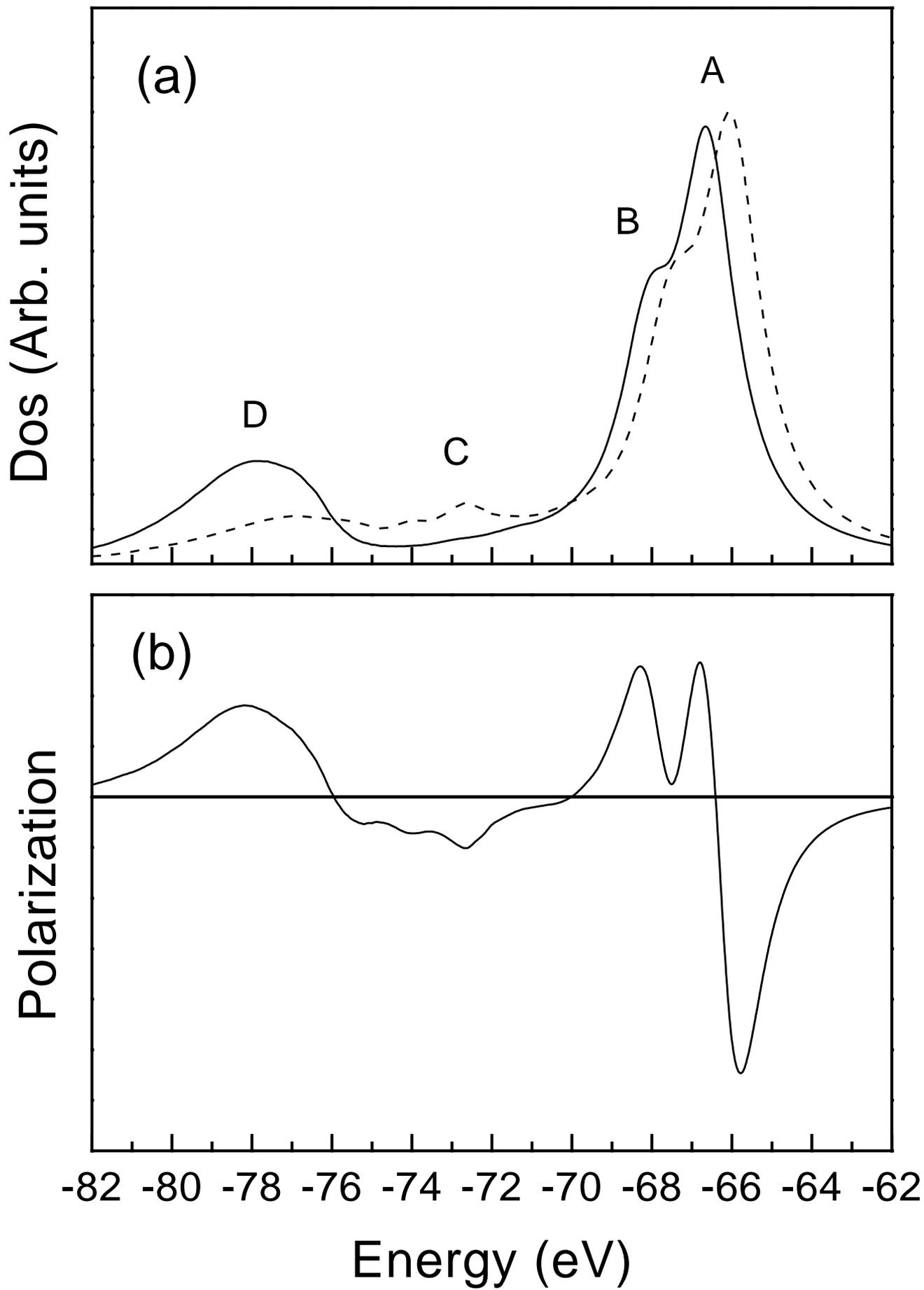
\end{document}